# Leveraging Fast Healthcare Interoperability Resources (FHIR) for the Exchange of Wearable Inertial Sensor Data: Design and Development Study

Gabriela Ciortuz[1,2], MSc; Joshua Wiedekopf[2], MSc; Hannes Ulrich[3], Dr rer nat; Hawzhin Hozhabr Pour[1], Dr-Ing; Sebastian Fudickar[2,4], Dr rer nat habil

[1]Institute of Medical Informatics, University of Luebeck, Lübeck, Germany

[2]Section for Clinical Research IT, Institute of Medical Biometry and Statistics, University of Luebeck & University Hospital Schleswig-Holstein, Lübeck, Germany

[3]Department of Medical Informatics, Institute for Community Medicine, University Medicine Greifswald, Greifswald, Germany

[4]Fraunhofer Research Institution for Individualized Medical Technology and Engineering (IMTE), Lübeck, Germany

**Corresponding Author:**

Gabriela Ciortuz, MSc

Institute of Medical Informatics

University of Luebeck

Ratzeburger Allee 160

Lübeck, 23562

Germany

Phone: +49 451 3101 5643

Email: gabriela.ciortuz@uni-luebeck.de

# Abstract

**Background:** Continuous health and movement monitoring using wearables can improve assessment of disease progression including frailty and functional decline in older adults and help clinicians with tracking, diagnosis, and treatment of neurological movement disorders. However, raw sensor streams, in particular inertial measurement unit (IMU) sensor data, are typically stored in proprietary files that cannot be easily exchanged with or interpreted by electronic health records. Health Level Seven (HL7) Fast Healthcare Interoperability Resources (FHIR) standardizes communication between devices and IT systems in healthcare, but the existing FHIR profiles for personal health devices target spot measurements, such as blood pressure, and short waveforms, not longitudinal multichannel motion recordings together with their recording context and activity labels.

**Objective:** This study aimed to develop WHIR (Wearable Health Interoperable Recordings), a FHIR R5-based information model and implementation guide that captures the metadata and motion data acquired via wearable sensors, and to demonstrate its applicability in both machine learning research and home-based clinical assessment.

**Methods:** We derived metadata requirements from a review of 23 published human activity recognition datasets and formalized them in a FHIR-independent information model. The model was then mapped to FHIR R5; profiles, extensions, code systems, and value sets were authored in FHIR Shorthand and built with SUSHI. Raw sensor signals are kept in files referenced through *DocumentReference* resources, so that FHIR holds the metadata and activity labels. Terminology artifacts use SNOMED CT, LOINC, and ISO/IEEE 11073 concepts wherever available, with locally defined codes only where no external system covers the concept.

**Results:** The proposed implementation guide defines 9 profiles and 4 extensions, together with 10 code systems and 15 value sets. All profiles, extensions, and 67 example instances passed the conformance checks of the official HL7 FHIR IG Publisher without errors. Two use cases, machine learning training data and automated home-based geriatric assessment, were implemented as conformant example instances. We discuss how our model aligns with the HL7 Personal Health Device (PHD) Implementation Guide where applicable.

**Conclusions:** WHIR provides a FHIR R5-conformant representation of wearable IMU recordings together with their device metadata, sensor placement, and time-aligned activity labels, and covers concepts that the HL7 Personal Health Device Implementation Guide does not address. Conformant example instances demonstrate its applicability to machine learning research and home-based geriatric assessment; evaluation with complete datasets and independently implemented systems is needed to establish operational interoperability. The implementation guide and its sources are openly available for reuse and extension.

# 1. Introduction

Equipping patients and study participants with wearable sensors provides continuous insights into their physiological and functional state outside the clinic [1]. Unlike a clinical assessment at a single visit, a continuous recording shows how a patient moves and functions over days or weeks [2]. For this reason, wearables are increasingly used in clinical trials, where they are expected to provide more frequent and more objective measurements than clinic visits [3]. In healthcare, wearables can offer advantages over video-based monitoring. They can provide greater privacy protection and user acceptance, especially when integrated into consumer devices or small sensors attached to clothing, and are well suited to continuous assessment in everyday life [4]. Smartphones and body-worn sensors with built-in inertial measurement units (IMUs), which combine a triaxial accelerometer and gyroscope and may also include a magnetometer, have become common tools for capturing movement [5].

Human activity recognition (HAR) turns such movement data into information about what a person is doing by identifying physical activities from the sensor signals [6]. HAR is usually approached as a classification problem [4], and combining several sensors across the body improves recognition accuracy because it captures complex activities more completely [7]. Machine learning models trained on IMU data detect activities of daily living (ADLs) [6] and the execution of functional ability tests [8,9]. ADLs are a de facto clinical measure of functional status in older adults living independently at home [10]. Automatic ADL detection could show clinicians when an older adult's daily routine changes, long before the next appointment [1].

In clinical studies of technology-supported geriatric assessment, older adults wore a sensor belt during a standard assessment in a controlled laboratory setting and for 1 week in daily life [11]. The aim was to reduce the assessment to a minimal set of items covering mobility, strength, and balance that older adults can perform at home without supervision [9]. Assessment items that are part of everyday life can already be measured with a single waist-worn IMU, such as stair-climbing power as an indicator of lower-limb strength [8] and gait parameters such as step length [12].

At home, IMUs have been used to monitor rehabilitation, for example to classify shoulder rehabilitation exercises from a single sensor [5]. Beyond physical performance, gait reflects cognition. In mild cognitive impairment, gait slowing under a concurrent cognitive task is associated with progression to dementia [13].

An IMU recording can only be interpreted within its context: at which body part the sensor was worn, how it was configured and synchronized with other sensors, and which assessment the recording belongs to. Yet wearable recordings and their metadata rarely reach electronic health records (EHRs). This is one reason why wearable sensor data are not yet part of routine care [14]. For example, in monitoring in everyday life, data are recorded on personal devices without a clinician present, and transferring them into clinical systems requires the sending device and the receiving system to support the same interoperability standard [15]. Such a standard has to handle more than device metadata: a single recording session can produce hours of high-frequency multichannel samples, and the activity labels that make a recording usable for research or assessment refer to time intervals within it.

Health Level Seven (HL7) Fast Healthcare Interoperability Resources (FHIR) has been adopted both within and beyond hospital networks, including internet-based and home care systems [16]. In Germany, so-called Digitale Gesundheitsanwendungen (Digital Health Applications, DiGAs) are digital medical devices that can be prescribed and reimbursed by statutory health insurance. To be listed in the DiGA directory of the Federal Institute for Drugs and Medical Devices (BfArM), manufacturers must meet interoperability requirements, including the publication of applicable FHIR profiles for free use in a recognized FHIR registry [17]. Comparable requirements for wearables could ease their integration into clinical workflows [16].

The HL7 FHIR Personal Health Device (PHD) Implementation Guide (IG) is the most suitable existing FHIR specification that defines how measurements and device information from personal health devices, such as glucose meters, blood pressure monitors, and digital scales, can be represented using FHIR *Device* and *Observation* resources, based on the ISO/IEEE 11073 information model [18].

However, the PHD IG targets personal health devices that follow ISO/IEEE 11073-10206 or Bluetooth health device profiles [18], whereas consumer smartwatches and smartphones, which are increasingly used for longitudinal movement and health monitoring, often use different protocols and proprietary formats [16]. Their data remain in vendor platforms and rarely reach clinical workflows or research [1]. Our approach therefore does not depend on ISO/IEEE 11073 communication but reuses ISO/IEEE 11073 codes where they exist.

To our knowledge, no FHIR specification yet represents longitudinal, continuous multi-axis IMU recordings together with their sensor placement, configuration, and activity labels. In practice, each dataset stores IMU data in its own file layout and configuration format, as the 23 datasets reviewed in Section 2.1 show. Seneviratne et al [1] reach the same conclusion and call for domain-specific FHIR profiles and implementation guides for continuous sensor data.

We therefore propose WHIR (Wearable Health Interoperable Recordings), a FHIR R5 implementation guide that represents time-series motion data, links activity labels to the sensor streams they annotate, and encodes sensor placement, sampling frequency, and device metadata [19]. Its first version covers wearable IMUs, the most common sensors in HAR datasets, but its structure (device, configured measuring capability, sensor placement, referenced raw recording, and activity labels) is designed to extend to other wearable sensors. Because WHIR is expressed using standard FHIR resources rather than a proprietary schema, activity-recognition output can in principle be stored and queried alongside other clinical data on an R5-capable FHIR server.

This study aimed to develop and technically validate a FHIR R5-based information model and implementation guide for wearable IMU recordings, their device and recording metadata, and their activity annotations, and to demonstrate it in a machine-learning and a home-based clinical setting. The main contributions of this work are as follows:

- We derive the data and metadata requirements for describing wearable IMU recordings for healthcare use cases, by reviewing 23 HAR datasets.
- We propose an information model for wearable IMU recordings independently of FHIR and map it to FHIR R5 resources and profiles, taking existing specifications such as the PHD IG into account.
- We publish WHIR as an openly available FHIR R5 implementation guide whose profiles, extensions, and reusable code systems and value sets for sensor placement, attachment, device characteristics, and activity labels were validated with the HL7 FHIR IG Publisher.
- We demonstrate WHIR in two healthcare use cases, training data for machine-learning-based HAR and automated geriatric assessment at home, with conformant example instances.

WHIR thus enables standardized exchange of IMU recordings together with the metadata required to interpret them. If adopted, it could give care teams access to movement data that are currently difficult to obtain and allow researchers to reuse recordings across studies.

# 2. Methods

In this section, we describe the requirement analysis based on published HAR datasets and typical IMU recordings (Section 2.1), the information model derived from it and its mapping to FHIR R5 (Section 2.2), and the rules for profiling, extensions, and terminology (Section 2.3).

## 2.1 Requirement analysis

To determine which sensor configurations and data streams a standardized format must represent, we reviewed published HAR datasets and extracted, for each, the number, types, and configuration of its IMU sensors, their placement on the body, and the activities to be classified.

Shin et al [20] systematically reviewed HAR datasets across sensor modalities and derived a summary of the most widely used datasets. We started from this list and applied three inclusion criteria: (i) the dataset employs at least one body-worn IMU sensor (accelerometer, gyroscope, and/or magnetometer); (ii) the publication reports at least the type and body placement of each sensor; and (iii) the recognized activities are healthcare-relevant. We therefore excluded datasets aimed at driving safety or assembly-line monitoring. We supplemented this list with further healthcare-oriented HAR datasets from our own prior work and the related literature, retaining only those that met the criteria above.

Applying the inclusion criteria yielded 23 IMU-based HAR datasets. The included datasets cover the following application areas: ADLs and elderly-care monitoring; fall and clinical event detection; exercise and rehabilitation; and generic HAR benchmarks that include healthcare-relevant activities. Table S1 (Supplementary File 1) lists their sensor counts and types, sampling rates, sensor placements, application areas, and complete activity lists.

Figure 1 summarizes the 23 datasets by application area. The number of activity classes ranges from 2 (Daphnet) to 68 (CogAge 2, 61 atomic and 7 composite activities), the number of body-worn IMU devices from 1 to 12 (HHAR), the number of body placements from 1 to 9 (REALDISP), and the sampling rate from 20 Hz to 200 Hz. TUD was recorded at approximately 100 Hz but published downsampled to 2.5 Hz. Sensor placement varies by application area, motivating the placement value set and placement extensions described in Section 3.3.

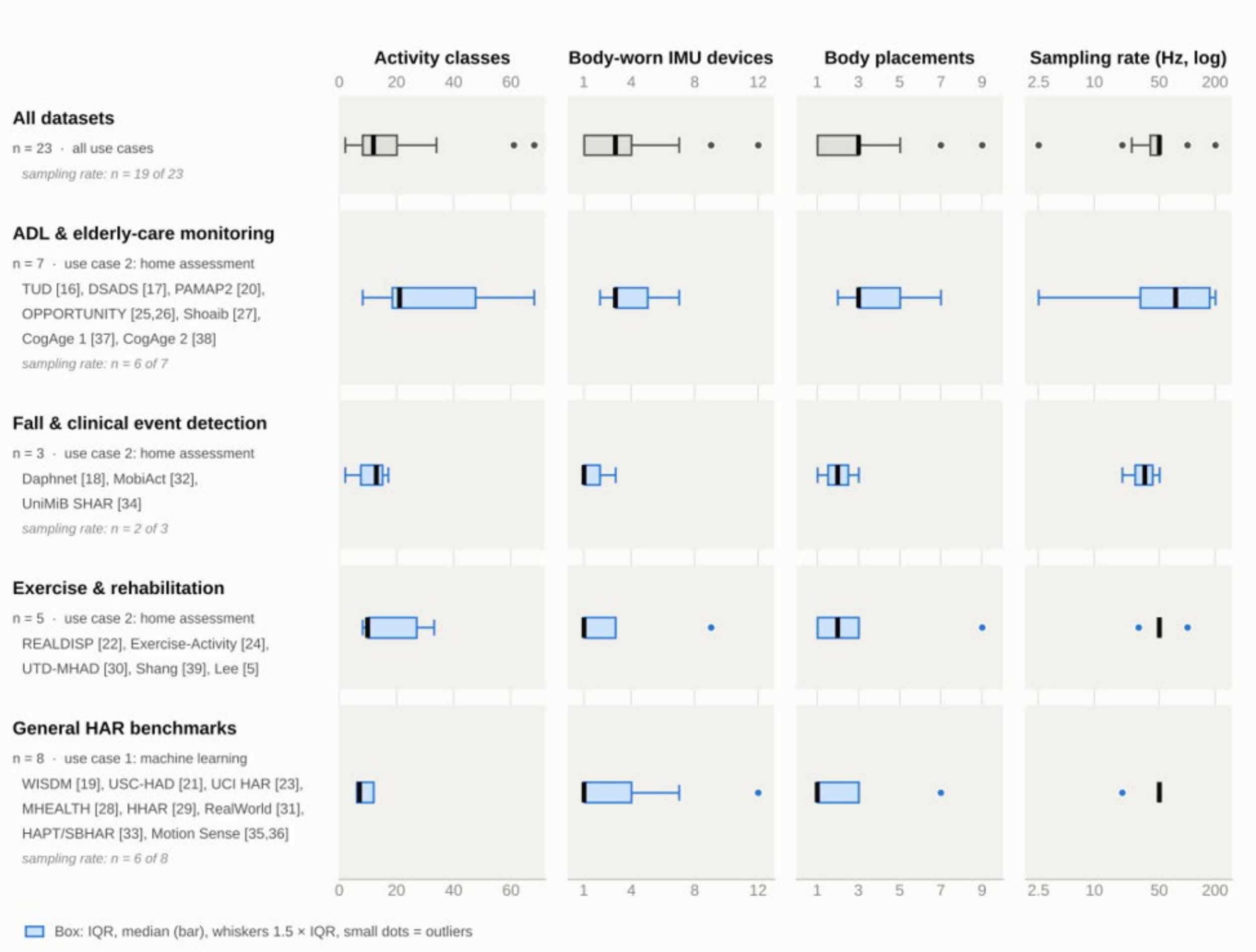

Figure 1. Characteristics of the 23 reviewed IMU-based HAR datasets by application area: number of activity classes, body-worn IMU devices, body placements, and sampling rate (log scale). Boxes show the interquartile range (IQR) with the median (bar); whiskers extend to 1.5 × IQR; dots are outliers. Datasets that do not report a sampling rate are omitted from that column.

To see how this information is stored in practice, we compared recordings from two of our own devices with those of two reviewed datasets (Table 1). Sampling rate and units appear in a configuration file, in column headers, in the file name, or only in the dataset documentation, and none of the four stores the sensor placement with the data.

Table 1. Sensors, sampling rate, data collection and output, and storage of activity labels for four wearable IMU devices: two devices from our own studies and the devices used in two public HAR datasets from the review. For PAMAP2 and UCI HAR, the output describes the published dataset.

| Device (source) | Sensors, ranges, and sampling rate | Data collection and output | Activity labels |
|---|---|---|---|
| Humotion SmarTracks sensor belt [21] (own recordings) | 3-axis accelerometer (±235 m/s²), gyroscope (±2000 °/s), and magnetometer (±400 μT); sampling rate configurable (250 Hz in the example file, Figure S1) | Recorded on the belt; one CSV file with all channels plus an INI configuration file with rates and units (Figure S2) | Must be documented in an external file |
| MbientLab MetaMotionS r1 [22] (own recordings) | 3-axis accelerometer (±2–16 g), gyroscope (±125–2000 °/s), and magnetometer; pressure, light, temperature; sampling rate configurable (50 Hz in our example file) | Logged on the device or streamed via Bluetooth LE; one CSV file per sensor, rate only in the file name, units in the column headers | Must be documented in an external file |
| Trivisio Colibri Wireless, 3 units (PAMAP2 [23]) | 3-axis accelerometers (±16 g and ±6 g), gyroscope, magnetometer, and temperature; 100 Hz (heart rate monitor ~9 Hz) | Transmitted wirelessly (2.4 GHz); published as one file per subject and session, rates and units only in the readme | Column in the data file |
| Samsung Galaxy S II (UCI HAR [24]) | Built-in 3-axis accelerometer and gyroscope; 50 Hz | Recorded on the phone worn at the waist; published as filtered 2.56 s windows, one file per signal and axis, rates and units only in the readme | Separate file, one label per window, labeled from video |

Based on the reviewed datasets, we consider an IMU recording to be adequately described by 10 metadata items: (1) start and end timestamp, (2) sensor identifier, (3) sensor type, (4) location and orientation of the sensor on the body, (5) sampling rate, (6) units of measurement, (7) calibration information, (8) subject information, (9) details of the data-acquisition encounter, and (10) the raw IMU data. Table 2 maps each item to FHIR R5 resources and elements. Several of these items are also needed to exchange wearable ECG data via FHIR [25]. In addition, the activity labels that annotate a recording must be linked to the time intervals of the raw data they refer to.

## 2.2 Information model and alignment with FHIR R5

From these requirements, we first developed an information model for wearable sensor recordings independently of FHIR (Figure 2).

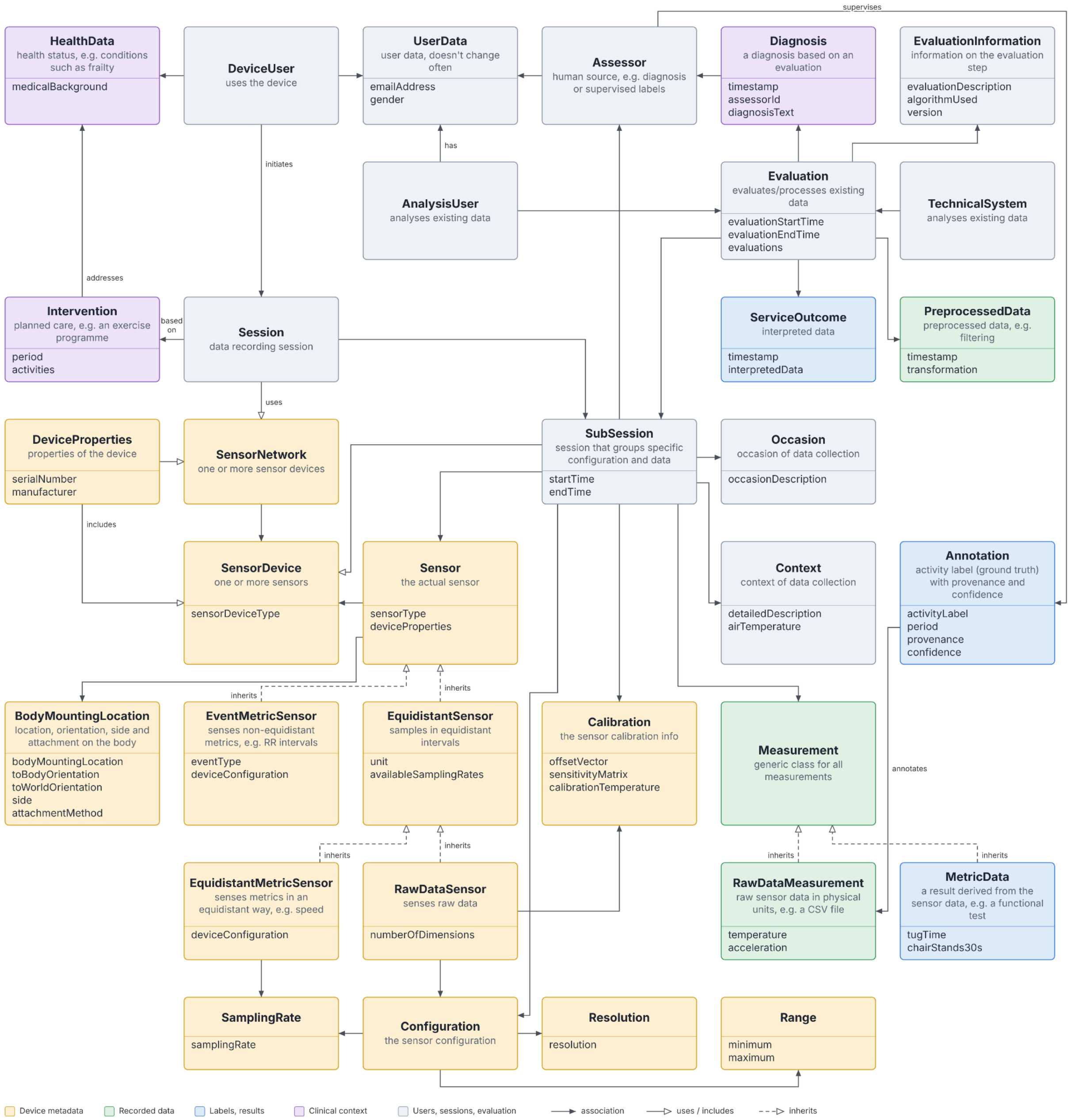

Figure 2. Information model for wearable sensor data.

We then mapped the information model to FHIR R5, published in March 2023 [26]; the choice of R5 over R4 is discussed in Section 4.3.3. The closest set of public FHIR profiles for our domain, the HL7 PHD IG, is based on FHIR R4 (v2.0.0, Standard for Trial Use [STU] 2). We therefore did not derive our profiles from it, but compared each resulting profile with its PHD counterpart where one exists (see Table 3). WHIR does not declare the PHD IG as a package dependency, because an R5 guide that depends on an R4 package produces validation errors. Instead, the three device identifier types taken from the PHD IG are defined in a local code system with the same codes (see Table S4).

Our FHIR mapping follows the FHIR R5 device model, which mirrors the ISO/IEEE 11073 domain information model (DIM): a medical device system (*Device*) has measurement capabilities (*DeviceMetric*) that produce observations (*Observation*).

For periodic signals, the PHD IG defines the *PhdRtsaObservation* profile, which embeds the samples in *Observation.valueSampledData*; since STU 2, multi-axis data such as x, y, and z acceleration can be reported as interleaved samples sharing one unit and scaling [18]. This suits waveforms of limited duration, such as ECG or spirometry traces. For continuous device streams, commercial FHIR conversion services group SampledData into one *Observation* per hour or per day [27]. Wearable-based HAR recordings, however, often span hours or days at sampling rates of 50-100 Hz or more, so even hourly segments contain hundreds of thousands of values per axis. We therefore developed our own observation profile for long IMU recordings. It references the raw data in an external file through a *DocumentReference*, so that FHIR resources hold only the metadata and the activity labels. It also requires references to the *DeviceMetric* describing the sensor configuration and the *BodyStructure* reference for the sensor placement.

We represented each activity label as a separate *Observation* that covers a time interval of the recording and references the raw data it annotates, so that labels remain individually queryable. In our examples, every observation carries its own identifier, and all observations of a recording session reference one *Encounter* whose identifier is the session identifier. Thus, a dataset can be split by session, for example into training and test set.

## 2.3 Profiling, extension, and terminology rules

We applied three rules when turning the information model into FHIR artifacts:

**Profiling.** We created a profile only when the base resource was too permissive for a requirement identified in Section 2.1 or in the use cases, for example, to make an optional reference mandatory or to restrict a choice element to one type. Otherwise, we reused the base resource without profiling.

**Extensions.** We added an extension only when no core FHIR R5 element captured the concept, and avoided immature R5 resources. *DeviceAssociation*, for example, could record where a device is worn, but it has maturity level 0 in R5 and is still under development, so we recorded placement through *Observation.bodyStructure* and the extensions on *BodyStructure* instead.

**Terminology.** Value sets and code systems were derived from the same review of the 23 published HAR datasets described in Section 2.1 and from our use-cases: for each characteristic requiring a coded value (body-structure placement, attachment mechanism, activity label) we recorded how each dataset described it, reused an existing SNOMED CT, LOINC, or ISO/IEEE 11073 concept wherever one existed, and defined a local code only where no external system covered the concept observed in the data.

The individual profiles, extensions, and value sets that result from applying these rules are presented in Tables 2-5. We validated all artifacts and example instances with the HL7 FHIR IG Publisher and demonstrated WHIR in 2 use cases (see Section 3.4).

# 3. Results

## 3.1 Mapping of the requirements to FHIR R5

Table 2 shows how each of the 10 metadata items identified in Section 2.1 is represented in FHIR R5. Sensor identity and type are held by the *IMUSensorDevice* profile; sampling rate, unit, and calibration status by *IMUDeviceMetric*; the sensor placement by a *BodyStructure* resource; and the raw data by a file referenced from the raw-data observation. Subject and encounter use the base *Patient* and *Encounter* resources. Sensor orientation can so far only be described as free text (see Section 4.4).

Table 2. Mapping of IMU recording metadata to FHIR R5 resources and elements.

| # | Metadata item | FHIR resource | Element |
|---|---|---|---|
| 1 | Start (and end) timestamp | *Observation* | *effectivePeriod.start* and *effectivePeriod.end* (first and last sample of the raw data file) |
| 2 | Sensor identifier | *Device* | *identifier* |
| 3 | Sensor type | *Device* | *type* and *conformsTo.specification* |
| 4 | Sensor placement on the body and orientation | *BodyStructure* | Alternative: *Observation.bodySite*. Orientation: free text in *BodyStructure.description*; no structured element yet (see Section 4.4) |
| 5 | Sampling rate | *DeviceMetric* | *measurementFrequency* |
| 6 | Units of measurement | *DeviceMetric* | *unit* |
| 7 | Calibration information | *DeviceMetric* | *calibration.type*, *calibration.state*, and *calibration.time* |
| 8 | Subject information | *Patient* | - |
| 9 | Data-acquisition encounter details | *Encounter* | *Observation.encounter*; *Encounter.identifier* holds the session identifier |
| 10 | IMU raw data | *DocumentReference* / Attachment, referenced from *Observation.derivedFrom* | *Observation.code* gives the sensor modality and *Observation.component* maps each axis to a file column. Alternatives: |

*Observation.valueSampledData* or *Observation.valueAttachment* (see Section 4.3.2). File format: minimum CSV format defined in the IG (code whir-csv).

## 3.2 FHIR-based model for motion sensor data

By mapping the information model (Figure 2) to FHIR R5, we derived the model shown in Figure 3. It distinguishes 5 groups of resources: static device metadata (*Device*, *DeviceDefinition*), which do not change between recordings; session-specific data (raw-data observations, document references, encounters); observations generated by machine-learning or annotation software; clinical context explaining why the data were recorded (eg, *CarePlan*, *Condition*, *DiagnosticReport*); and externally defined profiles that WHIR does not constrain further.

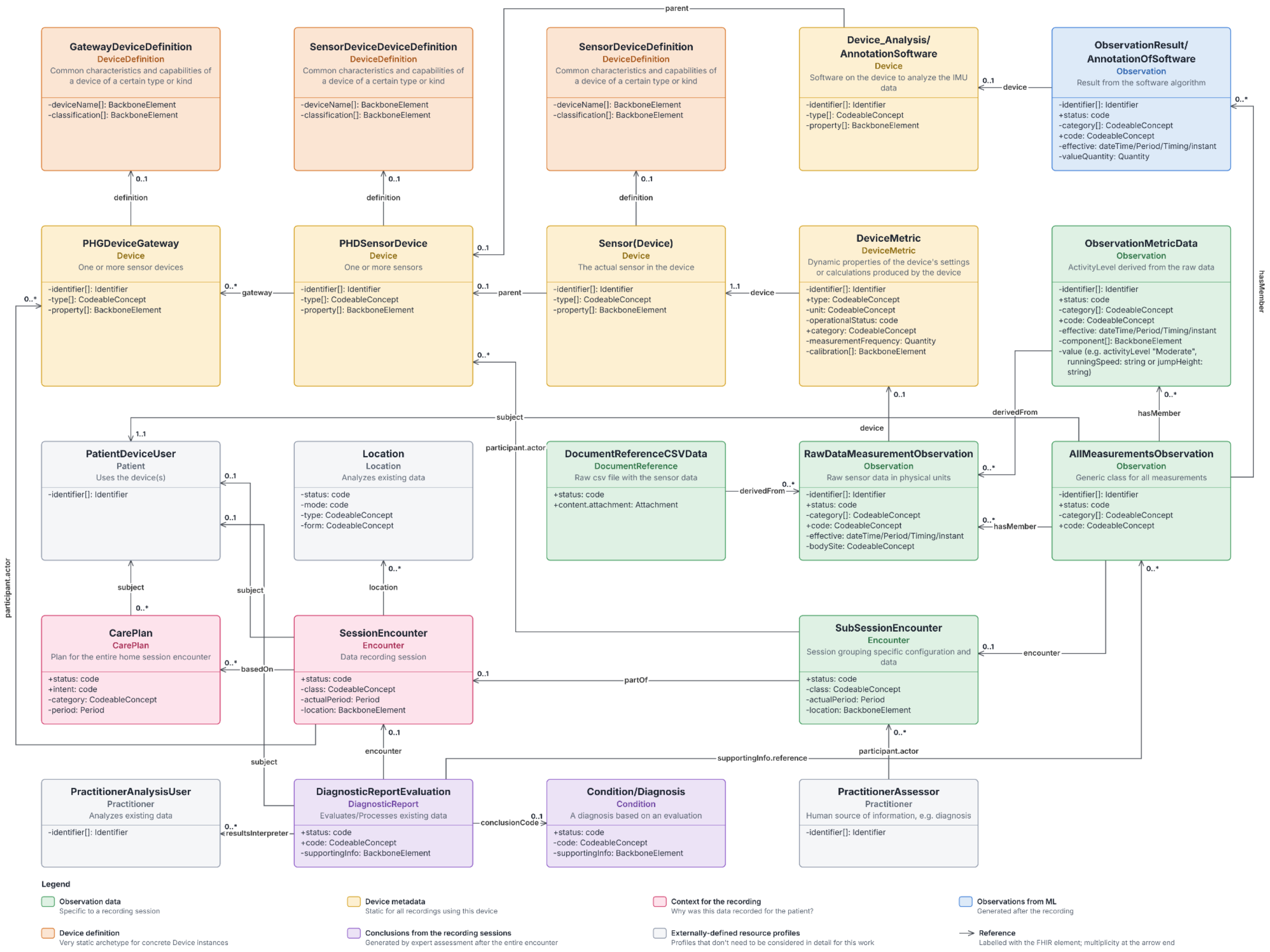

Figure 3. FHIR R5-based model for motion sensor data.

## 3.3 Profiles, extensions, and terminology artifacts

The model of Section 3.2 is implemented as a public FHIR R5 Implementation Guide (WHIR, package fudickar-lab.whir, built with SUSHI/FSH) [19]. The IG defines 9 profiles and 4 extensions (Table 3, Table 4), together with 10 code systems and 15 value sets (Table S4, Supplementary File 2). Figure 4 shows the profiles and the references between them.

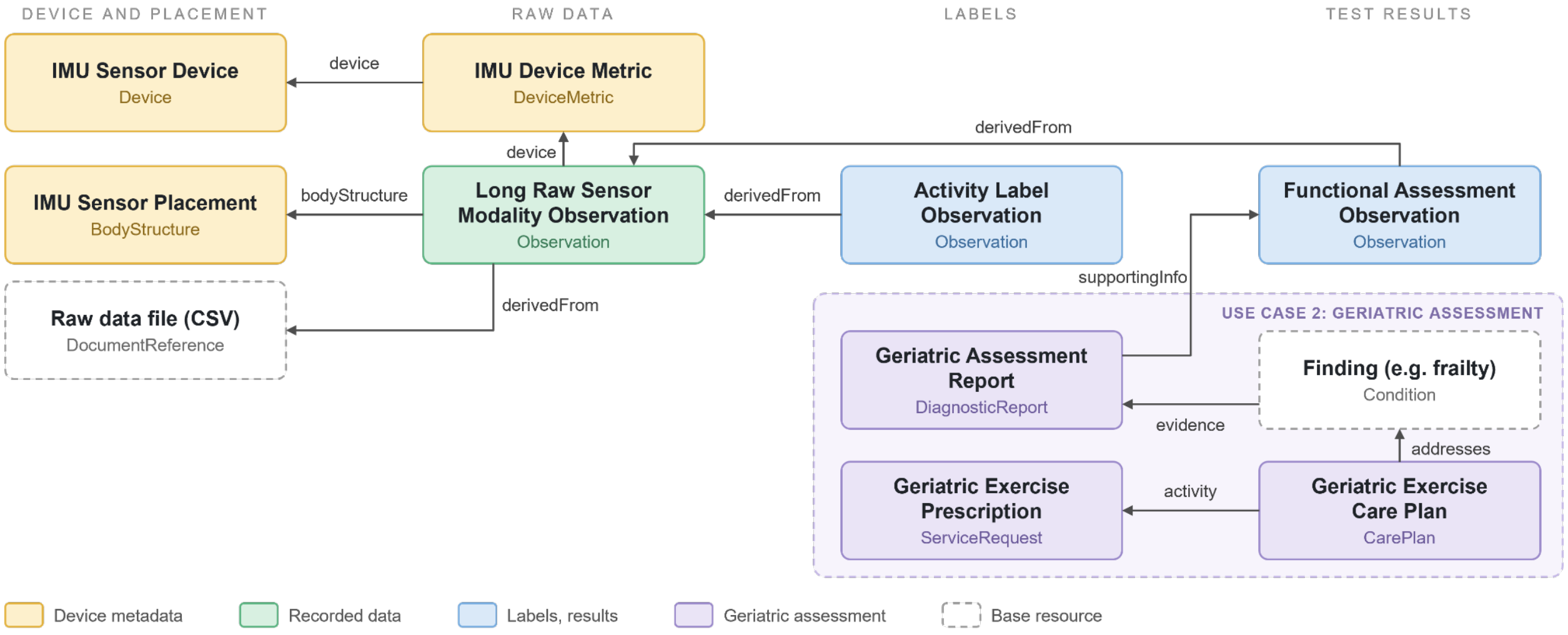

Figure 4. Profiles of the WHIR implementation guide and the references between them.

Each *LongRawSensorModalityObservation* holds one sensor modality of one device, for example the triaxial accelerometer of a sensor belt. Its *code* gives the modality (SensorModality value set), and one component per axis names the column of the raw data file that holds that axis (SensorAxis value set). The observation carries no *value[x]*: the samples are in the file referenced through *derivedFrom* (a *DocumentReference*, in our examples with content type text/csv). Sampling rate, unit, and calibration are given by the referenced *IMUDeviceMetric*, and the sensor placement by the *BodyStructure* reference. A device with an accelerometer, a gyroscope, and a magnetometer is therefore represented by three observations, each referencing its own *DeviceMetric*.

The *effectivePeriod* of a raw-data observation runs from the first to the last sample of its file. The IG defines a minimum format for this file: a UTF-8 CSV file with one header row, the Unix epoch timestamp in milliseconds in the first column, and one column per axis, named as in *Observation.component* and given in the unit of the *DeviceMetric*. The samples that belong to a label are those whose timestamps lie within the label's *effectivePeriod*, which must itself lie within the *effectivePeriod* of the raw-data observations it is derived from.

Three value sets are connected: ActivityLabels combines HARActivityLabels, GeriatricAssessmentLabels, and ADLLabels, so that *ActivityLabelObservation.code* can be bound to a single value set. ADLLabels is kept separate because *GeriatricExercisePrescription* binds only GeriatricAssessmentLabels: an exercise program can prescribe a chair-stand exercise, but not an everyday activity such as washing dishes. AnnotationProvenance is bound with required strength by *ObservationAnnotationProvenance*, because its 4 codes (human expert, human reviewed, semi-automated, algorithm-generated) form a closed list.

We record this with an extension rather than a FHIR *Provenance* resource, because one *Provenance* resource per label would double the number of resources and require an additional query to filter labels by how they were produced. *Provenance* is therefore recorded per processing run rather than per label: one *Provenance* resource lists the observations a run produced, the raw-data observations it read, the people involved, and the software *Device*, whose version identifies the software and model used. Keeping labels as separate observations also supports parallel annotation sets (eg, “expert annotation” vs “algorithm-generated label”), distinguished by *Observation.code* or *Observation.category*: the raw data stay unchanged while each label set lives in its own set of observations.

AttachmentMethod's codes were derived by inspecting the attachment mechanism reported in the publication of each of the 23 datasets listed in Table S1 (Supplementary File 1): 13 datasets name a mechanism the value set covers for all their sensor placements; 3 datasets (TUD [28], OPPORTUNITY [29,30], Shoaib [31]) name a mechanism for some but not all of their sensor placements; and 7 datasets (DSADS [32], Daphnet [33], PAMAP2 [23], REALDISP [34], UTD-MHAD [35], HAPT/SBHAR [36], Shang [37]) never state how the sensor is attached at all. Table S3 lists the resulting codes together with the datasets that motivated each.

Table 3. Description of our proposed profiles and, if applicable, comparison with the closest PHD IG profiles. All our profiles derive from the base FHIR R5 resources, not from PHD IG profiles.

| WHIR profile | Closest PHD profile or base resource | Difference | Comment |
|---|---|---|---|
| *LongRawSensorModalityObservation* | *PhdRtsaObservation* | Represents a periodic sample series like *PhdRtsaObservation*, but covers all axes of a sensor modality in one observation and references the samples in an external file (*derivedFrom → DocumentReference*) instead of embedding them in *valueSampledData*; sampling frequency, unit, and calibration are described by the referenced *IMUDeviceMetric*. Adds required *device* (→ *DeviceMetric*) and *bodyStructure* references. | The bodySite element used by the PHD profile is replaced in our profile by the mutually exclusive R5 *bodyStructure* reference. The referenced *BodyStructure* carries extensions for the attachment method and placement qualifiers such as the dominant side. |
| *IMUSensorDevice* | *PhdDevice* | Our profile does not need the *PhdDevice* specialization extension: in FHIR R5 this is covered by conformsTo.specification. It requires the type of an *IMUSensorDevice* to be one of: Simple MDS, Smartphone, or generic Device (physical object). | FHIR R5 supports device hierarchies via *Device.parent*, allowing different sensor modalities to be represented as parts of an IMU device or smartphone. |
| *IMUDeviceMetric* | *PhdDeviceMetric* | FHIR R4 already allows *Observation.device* to reference a *DeviceMetric*, but the PHD IG restricts it to Reference(*PhdDevice*), ie, the hardware rather than a specific configured measuring capability. Our *LongRawSensorModalityObservation* references the *IMUDeviceMetric* instead. In R5, *DeviceMetric.source* was renamed to *DeviceMetric.device*, which our *IMUDeviceMetric* adopts. | |
| *IMUSensorPlacement* | *BodyStructure* (base resource; no PHD equivalent) | The PHD IG defines no profile on *BodyStructure* and expresses placement, where it does so at all, through *Observation.bodySite*. | Binds the placement to the HARBodyLocations value set (extensible) and adds the optional extensions *BodyStructurePlacementQualifier* and *BodyStructureAttachmentMethod*. |
| *ActivityLabelObservation* | *Observation* (base resource; no PHD equivalent) | The PHD IG has no equivalent: it defines device measurements, not human-activity ground-truth labels. | Constrains effective[x] to Period (a label always covers a time span, never an instant); requires |
| | | | derivedFrom, restricted to *LongRawSensorModalityObservation*, and identifier (1..*); binds *code* to ActivityLabels (extensible, Section 3.3); *device* names the annotator or annotation software. |
| *FunctionalAssessmentObservation* | *PhdNumericObservation* | Records the result of a standardized functional test (code from FunctionalAssessmentTests, eg, the TUG time in seconds) that is computed from wearable sensor data or measured by an assessor, rather than a value measured directly by a personal health device. | Requires derivedFrom (1..*), restricted to *LongRawSensorModalityObservation*, as well as *encounter*, *effective[x]*, and *value[x]*; a quantity value must carry a UCUM unit code. The effective period must lie within the effective period of the raw-data observations it is derived from. *device* names the analysis software; the annotation provenance and confidence extensions are the same as for *ActivityLabelObservation*. |
| *GeriatricAssessmentReport* | *DiagnosticReport* (base resource; no PHD equivalent) | The PHD IG has no equivalent: it defines no reporting/interpretation resource for a set of related measurements. | Requires *supportingInfo* (1..*) to reference the *FunctionalAssessmentObservation* instances it interprets (eg, a Timed Up and Go or 30-second chair stand result) rather than *result*, because these measurements are captured during the recording and only interpreted, not produced, by the report. *conclusionCode* stays a CodeableConcept, because R5 *DiagnosticReport* has no element that references a *Condition*; the link runs the other way, through *Condition.evidence*. |
| *GeriatricExerciseCarePlan* | *CarePlan* (base resource; no PHD equivalent) | The PHD IG has no equivalent. FHIR R5 removed *CarePlan.activity.detail* [38], the inline element R4 used to code a planned activity directly on the *CarePlan*; a | Requires each activity to reference a *GeriatricExercisePrescription* instead of coding the exercise on |
| | | planned activity's code now has to live on whichever resource activity.plannedActivityReference points to. | *CarePlan* itself; *period* is required, because the program's duration is clinically meaningful (eg, a 2-week course of Otago exercises, a home-based strength and balance program for fall prevention [39]). |
| *GeriatricExercisePrescription* | *ServiceRequest* (base resource; no PHD equivalent) | The PHD IG has no equivalent. R5 *CarePlan* can no longer carry the activity code itself (see above); *GeriatricExerciseCarePlan.activity.plannedActivityReference* points to it. | *code.concept* is bound to *GeriatricAssessmentLabels* (extensible), the same locally defined Otago/rehabilitation vocabulary used to label recorded activity, so what was prescribed and what was recorded use identical codes; *occurrence[x]* (Timing) optionally states frequency and duration, eg, "3x/week for 2 weeks". |

Table 4. Extensions defined by the proposed model.

| Extension | Resource | Definition | Comment |
|---|---|---|---|
| *BodyStructurePlacementQualifier* | *BodyStructure* | *valueCodeableConcept* from PlacementQualifier (extensible) | States which structure is meant when a body structure without a side is referenced. |
| *BodyStructureAttachmentMethod* | *BodyStructure* | *valueCodeableConcept* from AttachmentMethod (extensible) | Records how the device was fixed to the body at this placement. |
| *ObservationAnnotationProvenance* | *Observation* | *valueCodeableConcept* from AnnotationProvenance (required) | States how a label was produced: by a human annotator, by a human reviewing/correcting an algorithm's output, by a combined workflow, or by an algorithm alone. |
| *ObservationAnnotationConfidence* | *Observation* | *valueDecimal*, 0.0–1.0 | The producing algorithm's confidence in the label; only meaningful alongside an algorithm-generated or semi-automated *ObservationAnnotationProvenance*. |

Table 5. Codes populating the PlacementQualifier value set, with the dataset placements that motivated each, bound by the *BodyStructurePlacementQualifier* extension (Table 4).

| Placement qualifier and datasets | System | Code | Display | Comment |
|---|---|---|---|---|
| Dominant ankle, ankle of the dominant leg [23] | http://snomed.info/sct | 262379005 | Dominant side (qualifier value) | The sensor is placed on the participant's dominant limb or side, for example the dominant wrist. |
| — | http://snomed.info/sct | 262458006 | Non-dominant side (qualifier value) | Non-dominant. The sensor is placed on the participant's non-dominant limb or side. |
| Placed by the subject as they preferred [24] | https://fudickar-lab.github.io/WHIR/CodeSystem/placement-qualifier-cs | preferred | Preferred | The participant or protocol freely chose this placement, without anatomical dominance being the criterion. No single-concept SNOMED CT equivalent, so defined locally. |
| Lateral side of the affected arm [5] | https://fudickar-lab.github.io/WHIR/CodeSystem/placement-qualifier-cs | affected | Affected | The sensor is placed on the clinically affected side, for example post-stroke or post-surgical. No single-concept SNOMED CT equivalent, so defined locally. |
| Executions with either the left or right hand were distinguished, to label whether the arm carrying the sensor performed the activity: LHO (Left Hand Only) [40] | http://snomed.info/sct | 255208005 | Ipsilateral (qualifier value) | The sensor is on the same side as the limb performing the recorded activity, for example sensor on the left arm, activity performed with the left hand. |
| — | http://snomed.info/sct | 255209002 | Contralateral (qualifier value) | The sensor is on the opposite side from the limb performing the recorded activity, for example sensor on the left arm, activity performed with the right hand. |

The corresponding SNOMED CT body-structure concepts for sensor placement, with notes on specificity and coverage, are given in Table S2 (Supplementary File 2).

## 3.4 Validation

We built WHIR with SUSHI v3.20.1 and validated all 9 profiles, 4 extensions, 10 code systems, 15 value sets, and 67 example instances (31 for use case 1 and 36 for use case 2) with the HL7 FHIR IG Publisher v3.0.2 against FHIR R5 (5.0.0). The IG Publisher checks every artifact and example instance against the FHIR R5 specification and the WHIR profiles, checks coded values against their bound value sets using the HL7 terminology server, and reports each issue as an error, a warning, or information. We considered the implementation guide conformant if the build reported no errors, and we reviewed all warnings individually. The build reports no errors, 131 warnings, and 146 information messages. None of the warnings concerns conformance of the example instances to the WHIR profiles; they relate to publication metadata (value sets marked as experimental, missing OIDs, and the *CodeSystem.caseSensitive* element), best-practice recommendations such as that every *Observation* should have a *performer*, and example URLs that do not resolve. In addition, we demonstrated WHIR in 2 use cases that place different demands on the model: supervised laboratory recordings with several devices, used as training data for machine learning, and unsupervised assessment at home, in which sensor-derived measures enter clinical documentation. For each use case, we authored FHIR Shorthand instances that represent an example recording session, from the patient, devices, and device metrics through the raw-data observations and referenced raw data files to the activity labels or functional test results and, in use case 2, the assessment report, condition, and care plan derived from them.

### 3.4.1 Use case 1: IMU data for machine learning

In this use case, IMU data of subjects are recorded in a controlled laboratory setting to develop machine-learning models for HAR, fall detection, or the automation of geriatric assessments. Multiple IMU devices, for example, a MetaMotionS (MMS) sensor [22], and a smartphone, are used simultaneously to capture a comprehensive set of sensor data, so this use case focuses on the metadata of the different IMU sensors, their positioning, and the activity labels for supervised machine learning. The sensor placements are encoded with the value set of Table S2 (Supplementary File 2): SNOMED CT 263301002, “Level of the third lumbar vertebra”, for the MMS sensor worn on a belt (attachment method belt), and SNOMED CT 37117007, “Right inguinal region structure (body structure)”, for the smartphone carried in the right trouser pocket (attachment method pocket).

A practitioner (assessor) supervises the recording sessions to ensure accurate data collection and correct activity labeling. The assessor is represented as a *Practitioner* resource and referenced as performer (*Observation.performer*) of the activity-label observations whose labeling they supervise. If relevant, the study design could additionally be modeled with FHIR in the future.

The instance diagram in Figure 5 shows the resulting resources for an example recording session.

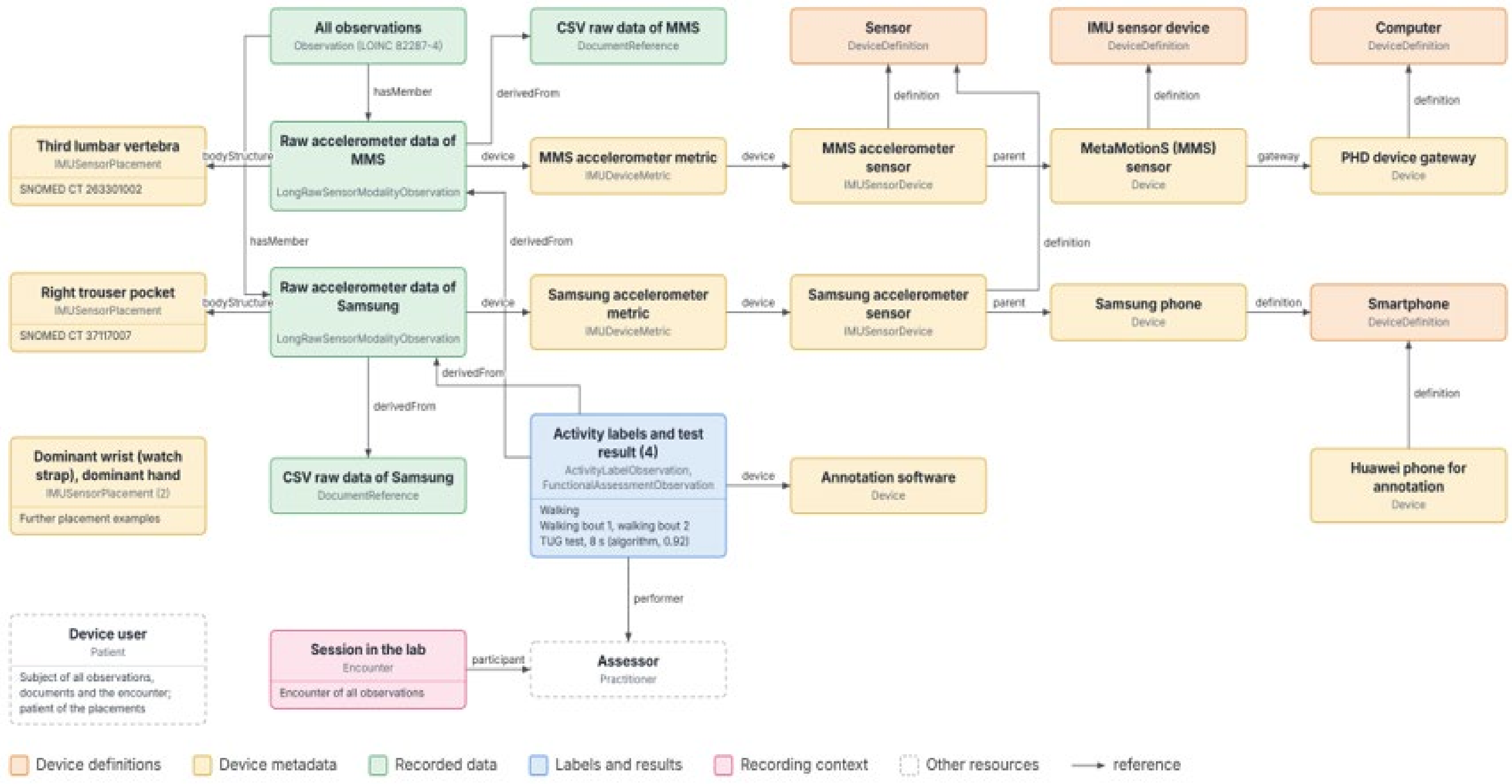

Figure 5. Instance diagram for use case 1: devices, device metrics, patient, raw-data observations with referenced CSV documents, body structures, and ML label observations.

### 3.4.2 Use case 2: Home assessment

In this scenario, automated geriatric assessments are performed in a home setting: the subject wears a minimal set of sensors (eg, one MMS sensor at the level of the third lumbar vertebra) and uses a mobile application that automatically generates the assessment results. The collected data are then transmitted to the practitioner for further analysis, or uploaded directly to the EHR for integration into the subject's medical record.

The recording session's raw accelerometer stream is analyzed by software that computes 2 standardized geriatric mobility measures, rather than the activity labels of use case 1: a Timed Up and Go result (LOINC 89423-8, "Time to rise from chair, walk 10 feet and back, and return to sitting [TUG]", reported as a Quantity in seconds) and a 30-second chair stand test result (LOINC 66247-8, "Sit to stand frequency in 30 seconds", reported as a Quantity in stands per 30 s). For patients who cannot complete the TUG, the timed Luebeck Scale of Basic Mobility (tLSBM) offers a graded alternative that rates basic mobility from lying to walking and times the transfers [41]; its result could be represented with the same *FunctionalAssessmentObservation* pattern.

Both *FunctionalAssessmentObservation* instances are referenced as *supportingInfo* of a *GeriatricAssessmentReport*, our profile on *DiagnosticReport*, coded as a "Geriatric medicine Outpatient Progress note" (LOINC 100467-0) and interpreted by the assessor, leading to a *Condition* capturing the geriatric conclusion (SNOMED CT 248279007 "Frailty (finding)", severity "Severe"). The *Condition* links back to the report through *Condition.evidence.reference* (*Condition.evidence* being a CodeableReference in R5, not R4's separate *evidence.detail*), since FHIR R5 gives *DiagnosticReport* no element that references a *Condition* directly. The recording itself is scoped to a home-health *Encounter* (*Encounter.class* = HH) at the patient's home (*Location*), based on a 2-week *GeriatricExerciseCarePlan* for a home exercise program. Because FHIR R5 removed *CarePlan.activity.detail*, each planned exercise is coded on a separate *GeriatricExercisePrescription* (our profile on *ServiceRequest*) that the *CarePlan* references through *activity.plannedActivityReference*, following the R5 *CarePlan* specification [38].

The instance diagram in Figure 6 shows the resulting resources for an example home assessment session.

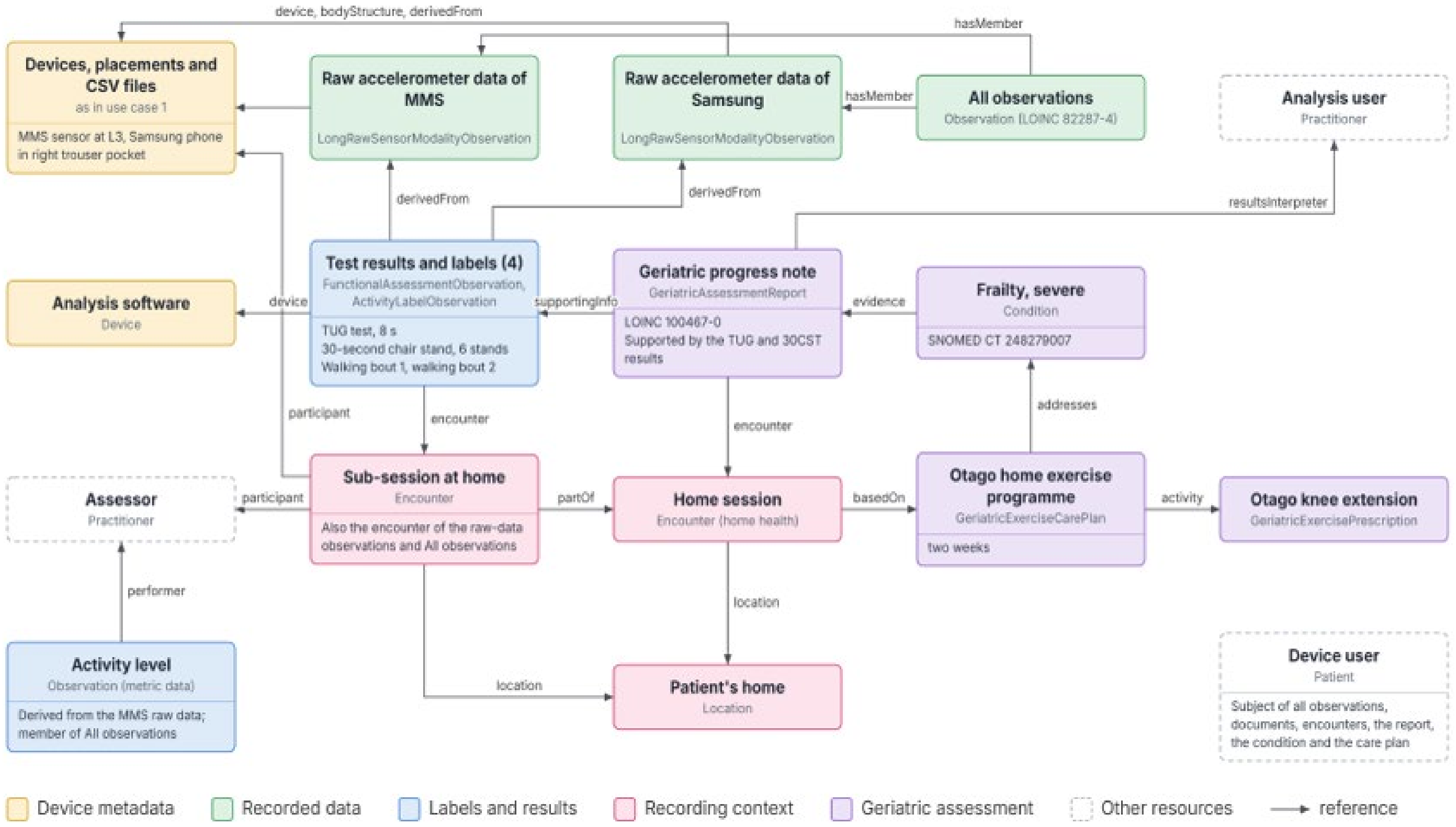

Figure 6. Instance diagram for use case 2: home-health encounter, devices, raw-data observations, TUG and 30CST *FunctionalAssessmentObservation* instances from analysis software, *GeriatricAssessmentReport*, *Condition*, and *GeriatricExerciseCarePlan*.

# 4. Discussion

## 4.1 Principal Results

This article proposes WHIR, a FHIR R5 information model and implementation guide for wearable IMU recordings. The implementation guide comprises 9 profiles, 4 extensions, 10 code systems, and 15 value sets, and 5 of the 9 profiles cover concepts for which the PHD IG has no counterpart. Validation of the implementation guide and 67 example instances using the HL7 FHIR IG Publisher demonstrated technical conformance to FHIR R5 and the WHIR profiles, without requiring changes to the core FHIR specification.

The two use cases indicate the technical feasibility of WHIR:

In use case 1, a supervised laboratory session with several devices, the representation provides what is needed to reuse a recording as machine-learning training data: which device recorded which channels, at what sampling rate, where on the body, and which labels apply to which time intervals.

In use case 2, an unsupervised assessment at home, the same raw-data observation profile is combined with functional test results that feed into a diagnostic report, a condition, and an exercise care plan, connecting sensor-derived measures to the clinical context in which they are used.

Three design decisions shaped the result. The information model was developed independently of FHIR and only then mapped to it, which keeps the model readable to domain experts and makes each mapping decision explicit. Raw signals are referenced through a *DocumentReference* rather than embedded in *Observation.valueSampledData*, allowing metadata and annotations to remain accessible as FHIR resources without embedding large signal files in the observations (see Section 4.3.2). Terminology reuses SNOMED CT, LOINC, and ISO/IEEE 11073 concepts wherever they exist, with locally defined codes only where no external system covers the concept.

## 4.2 Comparison with Prior Work

Prior work on FHIR for wearable data falls into three groups: application-specific adapters, specifications for activity summaries, and the PHD IG. We compare WHIR with them along three dimensions: how the raw recording is handled, how its recording context is described, and how activity annotations and derived clinical results are linked to it.

Several studies have built targeted FHIR adapters for wearable data. Bartschke et al [25] developed a Swift-based parser to convert Apple Watch ECG data into FHIR *Observation* resources for transmission to EHR systems. Gopinathan et al [42] extended FHIR to capture lifestyle data from smartphones. Pfyffer et al [43] implemented a standardized FHIR-based alerting system for fall detection using *Observation* and *Task* resources. Lee et al [5] classified shoulder-rehabilitation exercises from a single IMU and stored the results as FHIR *Observations*. Beyond device adapters, Leroux et al [44] mapped clinical study data from the CDISC Operational Data Model to FHIR, addressing both syntactic and semantic interoperability, and Gay and Leijdekkers [45] demonstrated mobile apps as enablers of interoperable connected health care. These studies demonstrate the use of FHIR for exchanging wearable-derived data in specific applications, typically focusing on processed results or events rather than the underlying recordings. Unlike WHIR, the approaches discussed here do not provide a common representation combining continuous IMU recordings, sensor placement, configuration, and time-aligned activity labels.

Two related standards address activity summaries. The HL7 FHIR US Physical Activity Implementation Guide (STU 1) standardizes the exchange of daily activity logs, step counts, heart-rate measures, and associated care plans and goals [46]. However, it explicitly excludes raw device data and information about the wearable sensors that produce such measures. IEEE 1752.1 (Open mHealth) standardizes episode-level sleep and physical-activity measures (activity type, duration, distance, intensity, and calories) through a JSON schema family that is independent of FHIR. Its device and provenance modeling is limited to a free-text "source_name" string and a binary "sensed/self-reported" modality flag in the header schema; it defines no structured device identity, sensor placement, or configuration metadata [47]. Both standards therefore focus on activity measures and summaries rather than the underlying continuous sensor recordings. WHIR complements these approaches by representing the raw IMU recordings and associated metadata from which such measures can be derived.

The PHD IG is the closest specification. Its *PhdRtsaObservation* profile can already carry a periodic sample series in one *Observation*, including multi-axis data since STU 2, and its device model, like WHIR, separates device identity, measurement capability, and observation [18]. However, *PhdRtsaObservation* embeds the samples in *Observation.valueSampledData*, which suits waveforms of limited duration, but not wearable HAR recordings that span hours or days at sampling rates of 100 Hz or more. WHIR therefore keeps the raw data in an external file and links it through a *DocumentReference*, so that FHIR resources hold only the metadata and the activity labels. The PHD IG does not allow such a link: in STU 2, *Observation.derivedFrom* is restricted to other PHD observations. WHIR also defines an explicit representation of all axes of a sensor modality in one observation, with one component per axis identifying the corresponding column in the raw-data file, whereas the PHD IG reports the axes only as positions in an interleaved sample array. Each raw-data observation references the configured measuring capability (*DeviceMetric*) and a structured sensor placement (*BodyStructure* with attachment method and placement qualifier). Activity labels are represented as separate observations linked to the recording, a concept not addressed by the PHD IG, which focuses on device measurements rather than activity annotations. Table 3 compares the individual profiles.

A further difference is in the representation of clinical results. The Smart Bear EU project's *TimedMobilityTestObservation* [48] models the Timed Up and Go test result as a categorical CodeableConcept rather than a raw time value, which does not preserve the original measurement precision. Use case 2 instead represents the measurement as a numeric LOINC-coded Quantity, preserving the measured value for potential longitudinal comparison across visits.

## 4.3 Design Considerations

### 4.3.1 Gateway architecture

In the PHD IG, the personal health gateway, typically the user's smartphone, receives ISO/IEEE 11073 data from the device, converts them to FHIR resources, and uploads them to a server, either immediately or after buffering [18]. WHIR deliberately leaves the location of this conversion open, because for IMU recordings the gateway's task differs in two respects. First, the raw signal itself is not converted: it remains a file, and the gateway only creates the metadata resources (*Device*, *DeviceMetric*, *BodyStructure*, the raw-data *Observation*, and its *DocumentReference*), which are known at the start of a recording and rarely change between sessions. Second, activity labels and functional test results are produced by annotation or analysis software that may run on the sensor, on the smartphone, or on a server, often after the recording has ended. WHIR therefore supports two deployment patterns. In an edge pattern, the smartphone app controlling the sensor creates the WHIR resources and uploads the raw file, so that a home assessment such as use case 2 reaches the EHR without a dedicated intermediary. In a server-side pattern, the raw stream is sent to a processing server that classifies activities and creates the label observations; in previous work, we streamed acceleration and gyroscope data from an Android app to such a server and classified activities in real time [49], and FHIR conversion could be added at that stage without changing the app. In both patterns, the annotation provenance extension (Table 4) and *Observation.device*, which names the annotation software, keep labels produced at different points of the pipeline distinguishable for the receiving system.

### 4.3.2 Storing raw data versus referencing it

An important design decision for wearable data is the choice between embedding the signal in *Observation.valueSampledData* and referencing an external file. FHIR supports regularly sampled physiological signals through SampledData, a fitting element in principle, because it documents the *interval*, *intervalUnit*, *lowerLimit* and *upperLimit*, and the dimensions of the data, all of which are relevant for longitudinal wearable signals. Embedding would keep the signal within a standardized FHIR representation, avoiding the need to parse an external CSV file and simplifying retrieval of the complete recording. This can suit very short signals, such as fall events or short clinically relevant excerpts. Wearable raw data, however, can involve long recordings: a 3-axis accelerometer sampled at 50 Hz over 24 hours produces 50 × 86,400 = 4.32 million samples per axis, that is, approximately 13 million values, or roughly 100 MB, in a single *Observation.valueSampledData* element. FHIR repositories are, moreover, not optimized to serve as time-series databases. Given the potential size of these recordings and the need to support different storage approaches, WHIR references the raw data rather than embedding them.

Referencing an external file can be done via *Observation.valueAttachment*, with the attachment as the observation value itself. However, because *Observation.value[x]* is a choice element, the same observation cannot then also carry the SampledData characteristics, losing the information about dimensions, intervals, and limits. In our design, the relevant recording metadata are instead represented by the referenced *IMUDeviceMetric* (sampling frequency, unit, calibration) and by the observation's components, which map each axis to a column of the raw data file.

WHIR links each raw-data observation to a *DocumentReference* through *Observation.derivedFrom*, allowing the raw signal to remain in an external file while its metadata and annotations are represented as FHIR resources. The raw file can also be stored as a FHIR *Binary* resource that the *DocumentReference* attachment points to. The FHIR specification allows the target of a *DocumentReference* to be a document on another server, a FHIR *Binary* resource, or a file on any other system [50]. For this approach, sender and receiver must agree on the file format, which WHIR specifies only as a minimum CSV format.

Referencing the raw data keeps the number of FHIR resources per recording small, but activity labels remain FHIR resources. Depending on the recording and annotation strategy, a HAR session may produce tens to hundreds of bout-level labels, potentially resulting in thousands of *ActivityLabelObservation* instances across a study. Retrieval scoped by subject, encounter, or *derivedFrom* relies on standard *Observation* search parameters, which servers commonly index, but we have not measured this in our setting (see Section 4.4). For cohort-wide analytics, such as aggregating label durations across a study, database-based approaches may be more efficient than individual FHIR REST queries, as suggested by published benchmarks [51,52]. FHIR's Bulk Data Access API supports cohort-scale export where the server implements it [53]. Local terminology caching may also improve ingestion performance when large numbers of labels require terminology validation [54].

### 4.3.3 FHIR version: R5 vs R4

We built WHIR directly on FHIR R5 rather than R4 (see Section 2.2), because no existing R4 deployment had to remain compatible with it. Löbe et al [55] faced a different situation when developing an extension module for the FHIR-based core dataset of the German Medical Informatics Initiative. Although R5 offered improvements for clinical-research modeling, their module had to remain compatible with existing R4-based modules, requiring selected R5 artifacts to be backported to R4. WHIR did not have this constraint.

The PHD IG is likewise based on FHIR R4. Its current version, v2.0.0 (STU 2), has no R5 counterpart. Choosing R5 allowed WHIR to use the newer resource definitions without maintaining compatibility with an existing R4 implementation, although it limits direct reuse of R4-based profiles such as those in the PHD IG.

## 4.4 Limitations and Future Work

WHIR has only been tested with illustrative example instances. Passing the IG Publisher's checks establishes technical conformance of the validated artifacts and example instances to FHIR R5 and the WHIR profiles, but does not demonstrate interoperability between independently developed systems, performance at realistic data volumes, or clinical utility. We have not yet mapped a complete dataset to it or measured server performance. Some concepts had no SNOMED CT or LOINC code, so we defined local codes. Finally, FHIR R5 is still less widely deployed than R4, and many clinical FHIR servers and EHR interfaces support only R4, which may slow adoption. Most R5 elements that WHIR uses have R4 counterparts (for example, *DeviceMetric.source* instead of *DeviceMetric.device*); the few without one, such as *Observation.bodyStructure* and *Device.conformsTo*, can be carried in R4 through the cross-version extensions published by HL7. If there is interest from implementers, we plan to provide an R4 version as future work.

Further development could extend the raw-data file format beyond the minimum CSV. The local codes could be proposed to SNOMED International and LOINC, and WHIR itself to the HL7 Devices Work Group for balloting as a Standard for Trial Use. Multimodal recordings are another direction. In mobile functional near-infrared spectroscopy (fNIRS), for example, a head-worn accelerometer measures the head movements that distort the brain signal [56]. Such recordings need a shared time base across devices, which WHIR covers only partly.

## 4.5 Conclusions

This design and development study presents WHIR, a FHIR R5 implementation guide for wearable IMU recordings comprising 9 profiles, 4 extensions, 10 code systems, and 15 value sets. Based on a review of 23 HAR datasets, WHIR provides a standardized representation that brings together device and sensor metadata, sensor placement, references to raw signals, and time-aligned activity annotations within FHIR. Conformant example instances for machine-learning-based HAR and home-based geriatric assessment demonstrate the technical applicability of this representation in research and clinical scenarios. As a next step, we will map complete datasets to WHIR and test exchange between independently implemented systems. WHIR is designed to extend to other wearable sensors, and the implementation guide is openly published for reuse and extension [19].

# Acknowledgements

We used OpenAI's ChatGPT to assist with language editing of the manuscript text and Anthropic's Claude to assist with refining the presentation of data reported in the manuscript.

# Funding

The study is funded by the German Federal Ministry of Research, Technology and Space (Project No. 01ZZ2007 and Project No. 01KX2626 in SV-NUKLEUS-NUM-MB and ENRICH). Apart from this public funding, the authors and their institutions did not receive payment or services from a third party for any aspect of the submitted work.

# Competing Interests

The authors have declared no competing interests. The authors and their institutions have not received payments or services from a third party in the past 36 months that could be perceived to influence, or give the appearance of potentially influencing, the submitted work.

# Data Availability

The WHIR implementation guide, including its FHIR Shorthand (FSH) sources and all example instances, is publicly available on GitHub [19]. The version described in this paper (version 0.1.0) is archived on Zenodo [68].

# Authors' Contributions

Conceptualization: GC (lead), SF, JW (supporting), HU (supporting)

Funding acquisition: SF

Investigation: GC (lead), HHP (supporting)

Methodology: GC (lead), JW (supporting), HU (supporting)

Resources: HHP (equal), SF (equal)

Software: GC

Supervision: SF

Validation: GC

Visualization: GC (lead), HHP (supporting)

Writing – original draft: GC

Writing – review & editing: GC (lead), JW (supporting), HU (supporting), HHP (supporting), SF (supporting)

# Ethics Approval

This study did not involve the recruitment of human participants, interventions, or the collection of new health data. The requirements were derived from published descriptions of publicly available datasets, and all FHIR example instances contain synthetic data. Ethics committee approval and informed consent were therefore not required. All relevant ethical guidelines have been followed.

# Abbreviations

**30CST:** 30-second chair stand test
**ADL:** activity of daily living
**API:** application programming interface
**BfArM:** Bundesinstitut für Arzneimittel und Medizinprodukte (German Federal Institute for Drugs and Medical Devices)
**CDISC:** Clinical Data Interchange Standards Consortium
**CSV:** comma-separated values
**DiGA:** Digitale Gesundheitsanwendung (digital health application)
**DIM:** domain information model
**ECG:** electrocardiogram
**EHR:** electronic health record
**FHIR:** Fast Healthcare Interoperability Resources
**fNIRS:** functional near-infrared spectroscopy
**FSH:** FHIR Shorthand
**HAR:** human activity recognition
**HL7:** Health Level Seven
**IEEE:** Institute of Electrical and Electronics Engineers
**IG:** implementation guide
**IMU:** inertial measurement unit
**IQR:** interquartile range
**ISO:** International Organization for Standardization
**JSON:** JavaScript Object Notation
**LOINC:** Logical Observation Identifiers Names and Codes
**MDS:** medical device system
**ML:** machine learning
**MMS:** MetaMotionS
**OID:** object identifier
**PHD:** Personal Health Device
**REST:** representational state transfer
**STU:** Standard for Trial Use
**SUSHI:** SUSHI Unshortens Short Hand Inputs
**tLSBM:** timed Luebeck Scale of Basic Mobility
**TUG:** Timed Up and Go
**UCUM:** Unified Code for Units of Measure
**WHIR:** Wearable Health Interoperable Recordings

# Supplementary Material

**Supplementary File 1**

IMU-based human activity recognition datasets reviewed (Table S1) and example raw IMU data and sensor configuration files (Figures S1 and S2).

**Supplementary File 2**

SNOMED CT body-structure concepts for sensor placement (Table S2), AttachmentMethod codes (Table S3), and code systems and value sets defined by WHIR (Table S4).

# Supplementary Material

## for: Leveraging Fast Healthcare Interoperability Resources (FHIR) for the Exchange of Wearable Inertial Sensor Data: Design and Development Study

Gabriela Ciortuz[1,2], MSc; Joshua Wiedekopf[2], MSc; Hannes Ulrich[3], Dr rer nat; Hawzhin Hozhabr Pour[1], Dr-Ing; Sebastian Fudickar[2,4], Dr rer nat habil

*Reference numbers in square brackets refer to the reference list of the main manuscript.*

## Supplementary File 1

Table S1. IMU-based HAR datasets relevant to healthcare reviewed in Section 2.1, with full activity lists, number and types of sensors, sampling rates, sensor placements, and application areas.

| Dataset | Activities (full list) | Sensors (count / type) | Sampling rate | Placement (body sites) | Application area |
|---|---|---|---|---|---|
| TUD [28] | Sitting/desk activities, Lying while reading/using computer, Having dinner, Walking freely, Driving car, Having lunch, Discussing at whiteboard, Attending a presentation, Driving bike, Watching a movie, Standing/talking on phone, Walking while carrying something, Walking, Picking up cafeteria food, Sitting/having a coffee, Queuing in line, Personal hygiene, Using the toilet, Fanning barbecue, Washing dishes, Kneeling/doing something else, Sitting/talking on phone, Kneeling/making fire for barbecue, Setting the table, Standing/having a coffee, Preparing food, Having breakfast, Brushing teeth, Standing/using the toilet, Standing/talking, Washing hands, Making coffee, Running, Wiping the whiteboard | 2 × Porcupine (3-D accelerometer, MCU, clock, 9 binary tilt switches, temperature sensor, 2 light sensors) | ≈ 100 Hz (raw); stored at 2.5 Hz for the published set | Right hip pocket; dominant (right) wrist | Elderly-care routine recognition; long-term activity monitoring |
| DSADS [32] | Sitting, Standing, Lying on back, Lying on right side, Ascending stairs, Descending stairs, Standing in elevator (still), Standing in elevator (moving), Walking in parking lot, Walking on treadmill (4 km/h, flat), Walking on treadmill (4 km/h, 15° incline), Running on treadmill (8 km/h), Exercising on stepper, Exercising on cross-trainer, Cycling on exercise bike (horizontal), Cycling on exercise bike (vertical), Rowing, Jumping, Playing basketball | 5 × Xsens MTx (each: 3-axis accelerometer + 3-axis gyroscope + 3-axis magnetometer) | 25 Hz | Both knees (sides), chest, both wrists | Remote monitoring of elderly, fall detection, home-rehab, multimodal vital-sign integration |
| Daphnet [33] | Freezing-of-Gait (FoG) detection during straight-line walking, FoG detection while turning, ADL tasks (fetch coffee, open doors, move between rooms) | 3 × accelerometers (body-worn) | – (device-specific) | Hip and leg (one sensor on hip, two on each leg) | FoG detection in Parkinson's disease |
| WISDM [57] | Walking, Jogging, Ascending stairs, Descending stairs, Sitting, Standing | 1 × smartphone accelerometer (≈ 20 samples / s) | ≈ 20 samples / s | Pocket (usually front) | Activity-based phone personalization; health-profile generation; large-scale activity monitoring |
| PAMAP2 [23] | Lying, Sitting, Standing, Walking, Running, Cycling, Nordic walking, Ironing, Vacuum cleaning, Rope-jump, Ascending stairs, Descending stairs, Watching TV, Using computer, Driving car, Playing soccer | 3 × IMUs (Colibri, Trivisio) + HR monitor | 100 Hz | Wrist (dominant arm), chest (central), ankle (dominant leg) | Physical-activity monitoring for aging populations; multimodal sensor research |
| USC-HAD [58] | 12 low-level ADLs (walk F/L/R, stairs up/down, run, jump, sit, stand, sleep, elevator up/down) | 1 × MotionNode (6-DOF IMU) | – (device-specific) | Front-right hip (pouch) | Benchmark for algorithm comparison; fitness & elder-care |
| REALDISP [34] | 33 fitness actions (walk, jog, run, various jumps, trunk twists, waist bends, arm lifts, rowing, cycling, …) | 9 × IMUs (accel + gyro + magnetometer) | 50 Hz | Trunk (back), left/right upper arm, left/right lower arm, left/right thigh, left/right calf | Detailed fitness analysis; multi-sensor HAR research |
| UCI HAR [24] | 6 ADLs (standing, sitting, lying, walking, walk-down, walk-up) | 1 × Samsung Galaxy S II (accel + gyro) | 50 Hz | Waist: left side of the belt (first trial); position chosen by the subject (second trial) | General-purpose activity-recognition baseline |
| Exercise-Activity [59] | 10 exercise actions (bench dips, squat, side-raises, shoulder press, curl, triceps extension, chest press, push-up, fly, bent-over row) | 1 × Nexus S 4G (accel + gyro) + 2 × MotoACTV (accel) | 30 Hz | Arm (armband), wrist (watch), hip (clip) | Health-fitness monitoring; home-exercise tracking |
| OPPORTUNITY [29,30] | 17 gestures (open/close drawer, dishwasher, fridge, door, cup, table, …) + 4 locomotion modes (stand, sit, walk, lie) + null | 72 sensors (24 × BLE accel/gyro, Sun SPOTs, InertiaCube3, Ubisense, magnetic sensor) | – (mixed) | 7 × body IMUs (right/left lower arm, right/left upper arm, back, right shoe, left shoe) + environmental sensors | Healthcare & assistive-technology research; gesture-based interaction |
| Shoaib [31] | Walking, running, sitting, standing, jogging, biking, stair-up, stair-down | 1 × Samsung Galaxy S II (accel, gyro, magnetometer, linear accel) | 50 Hz | Right/left jeans pockets; belt (right-leg side); right upper arm; right wrist | Healthcare & assisted-living monitoring |
| MHEALTH [60] | 12 activities (standing, sitting, lying, walking, stairs, waist bends, arm elevation, knee-crouch, cycling, jogging, running, jump) | Shimmer2 (accel + gyro + magnetometer) + chest ECG | 50 Hz | Chest, right wrist, left ankle (elastic straps) | Mobile-health benchmarking; multimodal body-sensing |
| HHAR [61] | Biking, sitting, standing, walking, stair-up, stair-down | 8 × smartphones + 4 × smartwatches | – (device-specific) | Phones in waist pouch; smartwatches on each arm | Large-scale activity-recognition benchmark |
| UTD-MHAD [35] | 27 actions (sports, hand gestures, daily activities, training exercises) | RGB video, depth video, skeleton, inertial sensor (50 Hz, ±8 g, ±1000 °/s) | 50 Hz | Right wrist, right thigh | Multimodal HAR; sports & rehabilitation |
| RealWorld [62] | Walking, jogging, stair-up/down, jumping, lying, standing, sitting, running | Smartphone (Galaxy S4) + MotoACTV smartwatch | 50 Hz | Chest, forearm, head, shin, thigh (pocket), upper arm, waist (various pockets/armbands/belts) | Position-aware activity recognition; on-body localization |
| Mobi-Act [63] | 9 ADLs + 4 fall types (forward-lying, front-knees-lying, sideward-lying, back-sitting-chair) | 1 × Samsung Galaxy S3 accelerometer | 20 Hz (5-s windows, 80 % overlap) | Free-form (any pocket or belt orientation) | Aging-well & e-health; fall detection |
| HAPT / SBHAR [36] | Real-time classification of standing, walking, sit-to-stand, sitting, walking-upstairs, sit-to-lie, etc. | 2 × accelerometers + 1 × gyroscope | 50 Hz | Smartphone on waist | Real-time health monitoring |
| UniMiB SHAR [64] | 9 ADLs + 8 fall types (various directions, protective strategies, syncope) | 1 × Samsung Galaxy Nexus (accelerometer) | 50 Hz | Front trouser pockets (left/right) | Fall-detection research; ADL monitoring |
| Motion Sense [65,66] | Downstairs, upstairs, walking, jogging, sitting, standing | iPhone 6s (accel, gravity, gyro, attitude), 12 channels via SensingKit | 50 Hz | Front pocket | General activity monitoring with mobile devices |
| CogAge 1 [40] | 61 atomic activities (6 pose states + 55 behavioral actions) | Smartphone (accel, gravity, gyro, linear accel, magnetometer @ 200 Hz) + Microsoft Band 2 (accel, gyro @ 67 Hz) + JINS MEME glasses (accel, gyro @ 20 Hz, blink & eye-movement) | 200 Hz (phone) / 67 Hz (band) / 20 Hz (glasses) | Smartphone front-left pocket; Band on left arm; Glasses on head | Detailed activity labeling for ML training; geriatric assessment |
| CogAge 2 | Same atomic set + 7 composite activities (brushing teeth, cleaning | LG G5 smartphone (accel, | 200 Hz | Smartphone front-left | Extended activity |
| [67] | room, handling medication, preparing food, styling hair, using phone, washing hands) | gravity, gyro, linear accel, magnetometer @ 200 Hz) + Huawei watch (accel, gyro @ 100 Hz) + JINS MEME glasses (accel @ 20 Hz) | (phone) / 100 Hz (watch) / 20 Hz (glasses) | pocket; Watch on left arm; Glasses on head | set for health-care & home-monitoring research |
| Lee 2024 [5] | 10 shoulder-rehab exercises (forward flexion, abduction, external rotation, internal rotation, scaption, horizontal abduction, horizontal adduction, extension, circumduction, rest) | 1 × commercial IMU (3-axis accel + 3-axis gyro), embedded in a smartphone or dedicated wearable | 50 Hz | Upper-arm (mid-humerus) strap | Automated classification of shoulder-pain rehabilitation exercises for home-based physiotherapy |
| Shang 2024 [37] | 8 Otago exercises (knee extension, knee flexion, hip abduction, hip extension, heel raise, toe raise, sit-to-stand, walking) | 1 × single-body-mounted IMU (3-axis accel + 3-axis gyro) | 100 Hz (down-sampled to 20 Hz for deep learning) | Waist (mid-line, just above the navel) | Home-based fall-prevention & geriatric rehabilitation; automatic adherence monitoring |

Sources: own prior work, CogAge 1 and 2; literature review, Daphnet, REALDISP, Lee 2024, and Shang 2024; all other datasets, Shin et al [20].

| Time-Date | Acc-X | Acc-Y | Acc-Z | Gyro-X | Gyro-Y | Gyro-Z | Mag-X | Mag-Y | Mag-Z | Power-Voltage |
|---|---|---|---|---|---|---|---|---|---|---|
| 1697535964064,00 | -0,08 | 0,936 | 7,664 | -0,073 | 0,007 | 0,023 | -25,870 | -5,410 | 56,150 | 4,016 |
| 1697535964068,00 | -0,09 | 1,070 | 8,460 | -0,073 | 0,007 | 0,023 | -25,780 | -5,689 | 56,058 | 4,016 |
| 1697535964072,00 | -0,099 | 1,146 | 8,930 | -0,073 | 0,007 | 0,023 | -25,690 | -5,968 | 55,966 | 4,016 |
| 1697535964076,00 | -0,099 | 1,194 | 9,208 | -0,073 | 0,007 | 0,023 | -25,960 | -5,593 | 55,686 | 4,016 |
| 1697535964080,00 | -0,109 | 1,213 | 9,381 | -0,073 | 0,007 | 0,023 | -26,230 | -5,219 | 55,405 | 4,016 |
| 1697535964084,00 | -0,099 | 1,222 | 9,467 | -0,073 | 0,007 | 0,023 | -25,870 | -4,948 | 55,680 | 4,016 |
| 1697535964088,00 | -0,099 | 1,232 | 9,525 | -0,073 | 0,007 | 0,023 | -25,510 | -4,677 | 55,955 | 4,016 |
| 1697535964092,00 | -0,071 | 1,251 | 9,563 | -0,108 | 0,061 | -0,061 | -25,690 | -4,951 | 55,772 | 4,016 |
| 1697535964096,00 | -0,071 | 1,251 | 9,573 | -0,001 | 0,002 | -0,004 | -25,870 | -5,225 | 55,589 | 4,016 |
| 1697535964100,00 | -0,08 | 1,251 | 9,554 | 0,004 | 0,001 | -0,002 | -25,600 | -5,045 | 55,958 | 4,016 |
| 1697535964104,00 | -0,08 | 1,251 | 9,544 | 0,003 | 0,001 | -0,001 | -25,330 | -4,865 | 56,328 | 4,016 |
| 1697535964108,00 | -0,08 | 1,261 | 9,544 | 0,003 | 0,004 | -0,002 | -25,690 | -4,674 | 56,143 | 4,016 |
| 1697535964112,00 | -0,08 | 1,261 | 9,544 | 0,001 | 0,005 | -0,002 | -26,050 | -4,483 | 55,958 | 4,016 |
| 1697535964116,00 | -0,09 | 1,261 | 9,544 | 0,001 | 0,006 | -0,002 | -26,140 | -5,313 | 55,778 | 4,016 |
| 1697535964120,00 | -0,08 | 1,270 | 9,544 | 0,001 | 0,006 | -0,001 | -26,230 | -6,143 | 55,599 | 4,016 |
| 1697535964124,00 | -0,09 | 1,261 | 9,544 | 0 | 0,005 | -0,002 | -26,050 | -5,407 | 55,778 | 4,016 |
| 1697535964128,00 | -0,08 | 1,251 | 9,554 | 0,001 | 0,005 | -0,001 | -25,870 | -4,671 | 55,958 | 4,016 |
| 1697535964132,00 | -0,09 | 1,261 | 9,563 | 0 | 0,005 | -0,002 | -25,780 | -4,672 | 55,770 | 4,016 |
| 1697535964136,00 | -0,099 | 1,261 | 9,554 | 0,001 | 0,005 | -0,002 | -25,690 | -4,674 | 55,583 | 4,016 |
| 1697535964140,00 | -0,09 | 1,261 | 9,563 | 0,003 | 0,002 | -0,001 | -25,960 | -5,039 | 55,309 | 4,016 |
| 1697535964144,00 | -0,099 | 1,261 | 9,573 | 0,004 | 0 | -0,001 | -26,230 | -5,404 | 55,034 | 4,016 |
| 1697535964148,00 | -0,099 | 1,261 | 9,573 | 0,004 | 0,001 | -0,002 | -26,230 | -5,034 | 55,124 | 4,016 |
| 1697535964152,00 | -0,099 | 1,261 | 9,563 | 0,005 | 0 | -0,001 | -26,230 | -4,664 | 55,215 | 4,016 |
| 1697535964156,00 | -0,09 | 1,270 | 9,563 | 0,004 | -0,002 | 0 | -26,050 | -4,852 | 55,494 | 4,016 |
| 1697535964160,00 | -0,09 | 1,270 | 9,563 | 0,002 | 0,001 | -0,002 | -25,870 | -5,040 | 55,774 | 4,016 |
| 1697535964164,00 | -0,08 | 1,270 | 9,563 | 0,003 | 0 | -0,001 | -25,780 | -4,765 | 55,585 | 4,016 |

Figure S1. Example of raw IMU data in CSV format, recorded at 250 Hz with a Humotion SmarTracks sensor belt: Unix timestamp (ms), triaxial accelerometer, gyroscope, and magnetometer, and supply voltage.

```
[Chan.10]
Type=Mag
ChanName=Y
ID=165
Freq=250.0
Sign=1
Unit=uT
Max_Value=400.000000
Min_Value=-400.000000

[Chan.11]
Type=Mag
ChanName=Z
ID=166
Freq=250.0
Sign=1
Unit=uT
Max_Value=400.000000
Min_Value=-400.000000

[Chan.12]
Type=Power
ChanName=Voltage
ID=155
Freq=250.0
Sign=0
Unit=V
Max_Value=6.000000

[Measure]
Start=2023-10-17T:10:46:04:064
Stop=2023-10-17T:12:20:47:064
DurationInMS=2083636
Duration=2083
```

```
[Ini]
Version=1
FloatValues=1
CustomFormat=2

[Channels]
Count=12

[Compression]
FrameSize=45
RecordsPerSecond=250.0
SuperFrameCount=1

[SuperFrame.1]
Freqs=250.0

[Chan.1]
Type=Time
ChanName=Date
ID=260
SecondsFromStartFactor=1000

[Chan.2]
Type=Acc
ChanName=X
ID=121
Freq=250.0
Sign=1
Unit=m/s^2
Max_Value=235.359604
Min_Value=-235.359604
```

Figure S2. Excerpts of the accompanying sensor configuration file: (a) file header and channel definitions; (b) further channel definitions and the measurement section, giving each channel's sampling frequency, unit, and value range.

# Supplementary File 2

Table S2. SNOMED CT body-structure concepts for sensor placements reported in the reviewed HAR datasets, with notes on specificity and coverage.

| Positioning and datasets | System | Code | Display | Comment |
|---|---|---|---|---|
| Chest (DSADS [32]), MHEALTH [60], RealWorld [62] | http://snomed.info/sct | 51185008 | Thoracic structure (body structure) | Synonym: Chest |
| Centrally on the chest [23] | http://snomed.info/sct | 279233000 | Structure of central surface region of chest (body structure) | Synonym: Central surface region of chest; More specific than the previous code, when dataset would specify that the sensor is centered on the chest (which is usually most common for HAR datasets). |
| Trunk (Back) (REALDISP [34]), Back (OPPORTUNITY [29,30]) | http://snomed.info/sct | 77568009 | Structure of posterior region of trunk (body structure) | Synonym: Back |
| Trunk (Back) (REALDISP [34]), Back (OPPORTUNITY [29,30]) | http://snomed.info/sct | 272612002 | Structure of surface region of back (body structure) | Synonym: Surface region of back; More specific than the previous code, because it specifies that the sensor is on the surface of the back, which is the case for IMU sensors. |
| Left side of the waist mounted with a belt, Placed by the subject as they preferred [24], Waist [37] | http://snomed.info/sct | 33673004 | Structure of waist (surface region) (body structure) | Synonym: Waist; doesn't denote the left side of the waist, users might think the sensor should be centered on the waist. |
| Hip [29,30,33,59] | http://snomed.info/sct | 29836001 | Hip region structure (body structure) | Synonym: Hip region; Side not specified. |
| Right hip pocket (TUD [28]), Front right hip [58] | http://snomed.info/sct | 287579007 | Right hip region structure (body structure) | Synonym: Right hip |
| Head (mounted with a belt, picture in the paper shows that the smartphone is placed on the left side of the head) (RealWorld [62]), glasses placed on the subjects' head [40], glasses worn [67] | http://snomed.info/sct | 69536005 | Head structure (body structure) | Synonym: Head; doesn't specify exactly where on the head. |
| Dominant wrist, wrist of the dominant arm [23,28], Wristwatch [59] | http://snomed.info/sct | 8205005 | Wrist region structure (body structure) | Synonym: Wrist; Side not specified; dominant side is not covered. |
| Dominant wrist, wrist of the dominant arm [23,28], Wristwatch [59] | http://snomed.info/sct | 244002004 | Structure of surface of wrist region (body structure) | Synonym: Surface region of wrist; Side not specified, more specific than the previous code, because it specifies that the sensor is on the surface of the wrist, which is the case for IMU sensors; dominant side is not covered. |
| Right wrist (DSADS [32], OPPORTUNITY [29,30], | http://snomed.info/ | 9736006 | Structure of right wrist region | Synonym: Right wrist; More specific than the |
| Shoaib [31], MHEALTH [60], UTD-MHAD [35]) | sct | | (body structure) | previous codes, because it specifies the right wrist. |
| Left wrist (DSADS [32], OPPORTUNITY [29,30], Left arm [40,67] | http://snomed.info/sct | 5951000 | Structure of left wrist region (body structure) | Synonym: Left wrist; More specific than the previous wrist codes, because it specifies the left wrist. |
| Upper-arm ( lateral side of the affected arm, aligned with the longitudinal axis of the humerus) | http://snomed.info/sct | 34894009 | Structure of lateral surface of upper arm (body structure) | Synonym: Lateral surface of upper arm; side is not specified; affected part is not covered. |
| Right upper arm (REALDISP [34], OPPORTUNITY [29,30], Shoaib [31], RealWorld [62], Right upper arm_, Right upper arm^ (OPPORTUNITY [29,30]) | http://snomed.info/sct | 368209003 | Right upper arm structure (body structure) | Synonym: Right upper arm. |
| Left upper arm (REALDISP [34], OPPORTUNITY [29,30], Left upper arm_, Left upper arm^ (OPPORTUNITY [29,30]) | http://snomed.info/sct | 368208006 | Left upper arm structure (body structure) | Synonym: Left upper arm. |
| Right lower arm (REALDISP [34]), OPPORTUNITY [29,30] | http://snomed.info/sct | 64262003 | Structure of right forearm (body structure) | Synonym: Right forearm; Some datasets use the term “right lower arm”. |
| Forearm (RealWorld [62]), Left lower arm (REALDISP [34], OPPORTUNITY [29,30]) | http://snomed.info/sct | 66480008 | Structure of left forearm (body structure) | Synonym: Left forearm; Some datasets use the term “Left lower arm”. |
| Placement of the smartphone is freely chosen (Mobi-Act [63]) | http://snomed.info/sct | 68367000 | Thigh structure (body structure) | Synonym: Thigh; used as the closest body site for a freely chosen phone position; the free placement itself is recorded with the AttachmentMethod code unconstrained-carry (Table S3). |
| Right thigh (REALDISP [34], UTD-MHAD [35]) | http://snomed.info/sct | 11207009 | Structure of right thigh (body structure) | Synonym: Right thigh. |
| Left thigh (REALDISP [34]), Thigh/Trouser pocket (RealWorld [62]) | http://snomed.info/sct | 61396006 | Structure of left thigh (body structure) | Synonym: Left thigh. |
| Right calf (REALDISP [34]) | http://snomed.info/sct | 722116009 | Structure of calf of right lower leg (body structure) | |
| Left calf (REALDISP [34]) | http://snomed.info/sct | 722115008 | Structure of calf of left lower leg (body structure) | |
| Shin (RealWorld [62]) | http://snomed.info/sct | 78234002 | Shin structure (body structure) | Synonym: Shin |
| Dominant ankle, ankle of the dominant leg [23] | http://snomed.info/sct | 244019002 | Structure of surface region of ankle (body structure) | Synonym: Surface region of ankle; side not specified; combine with the Dominant placement qualifier (Table 5). |
| Dominant ankle, ankle of the dominant leg [23] | http://snomed.info/sct | 6685009 | Structure of right ankle (body structure) | Synonym: Right ankle; use only if the dominant side is known to be the right side; otherwise use |
| | | | | 244019002 with the Dominant qualifier. |
| Left ankle (MHEALTH [60]) | http://snomed.info/sct | 51636004 | Structure of left ankle (body structure) | Synonym: Left ankle |
| Right hand [29,30] | http://snomed.info/sct | 78791008 | Structure of right hand (body structure) | Synonym: Right hand |
| Left hand [29,30] | http://snomed.info/sct | 85151006 | Structure of left hand (body structure) | Synonym: Left hand |
| On the sides of the knees, right side of the right knee (DSADS [32], Right knee^ (the image of the placement of the sensors shows the sensor mounted with a belt exactly over the knee), Right knee_ (under the knee) [29,30] | http://snomed.info/sct | 6757004 | Structure of right knee region (body structure) | Synonym: Right knee; doesn't denote the side of the knee, users might think the sensor should be centered on the knee. |
| On the sides of the knees, Left side of the left knee (DSADS [32]), Left knee^, Left knee_ [29,30] | http://snomed.info/sct | 82169009 | Structure of left knee region (body structure) | Synonym: Left knee; doesn't denote the side of the knee, users might think the sensor should be centered on the knee. |
| 3rd lumbar vertebra (UseCase1, UseCase2) | http://snomed.info/sct | 263301002 | Level of the third lumbar vertebra (body structure) | |
| Right hip pocket [28], Pocket (WISDM (Wireless Sensor Data Mining) Dataset [57] ), Right jeans pocket (Shoaib [31], Front trouser pocket (UniMiB SHAR [64]), Front pocket (Motion Sense [65,66]), (UseCase1, UseCase2) | http://snomed.info/sct | 37117007 | Right inguinal region structure (body structure) | Synonym: Right inguinal region |
| Trouser pocket (RealWorld [62]), Left jeans pocket (Shoaib), Front trouser pocket (UniMiB SHAR [64]), Front pocket (Motion Sense [65,66]), Left pocket of jeans [40], Left pocket [67] | http://snomed.info/sct | 85119005 | Left inguinal region structure (body structure) | Synonym: Left inguinal region |

Table S3 lists the codes populating the AttachmentMethod value set bound by the *BodyStructureAttachmentMethod* extension (Table 4). Only one concept could be reused from SNOMED CT, the general-purpose Strap (48990009); no external system covers the remaining mechanisms at the granularity the reviewed datasets report them, so 12 codes are defined locally in AttachmentMethodCS. Two of these, watch strap and elastic strap, are deliberately narrower than the reused SNOMED CT Strap concept, because the distinction between a watch-style fastening, an elastic band, and a general strap is reported consistently enough across datasets to be worth preserving.

Table S3. Codes populating the AttachmentMethod value set, with the dataset attachment mechanisms that motivated each, bound by the *BodyStructureAttachmentMethod* extension (Table 4).

| Attachment method and datasets | System | Code | Display | Definition |
|---|---|---|---|---|
| Right hip pocket [28], Pocket [57], Right/left jeans pocket [31], Front trouser pocket [64], Front pocket [65,66], Left pocket [40,67] | https://fudickar-lab.github.io/WHIR/CodeSystem/attachment-method-cs | pocket | Pocket | Device carried loose inside a garment pocket, not otherwise fixed to the body. |
| Belt [31], waist belt [24] | https://fudickar-lab.github.io/WHIR/CodeSystem/attachment-method-cs | belt | Belt | Device threaded onto, or clipped to, a waist belt. |
| Front-right hip pouch [58], waist pouch [61] | https://fudickar-lab.github.io/WHIR/CodeSystem/attachment-method-cs | pouch | Pouch | Device carried in a dedicated pouch worn on the body, distinct from a garment pocket. |
| Hip clip [59] | https://fudickar-lab.github.io/WHIR/CodeSystem/attachment-method-cs | clip | Clip | Device fastened directly to clothing with a clip mechanism, without a pocket, pouch, or belt. |
| JINS MEME glasses [40,67] | https://fudickar-lab.github.io/WHIR/CodeSystem/attachment-method-cs | frame-mounted | Frame-mounted | Device mounted on the frame of eyewear. |
| Right/left shoe [29,30] | https://fudickar-lab.github.io/WHIR/CodeSystem/attachment-method-cs | footwear-mounted | Footwear-mounted | Device built into or fixed to footwear. |
| Armband [59,62], Microsoft Band 2 on left arm [40] | https://fudickar-lab.github.io/WHIR/CodeSystem/attachment-method-cs | armband | Armband | Device secured to the upper arm with an armband. |
| MotoACTV smartwatch [62], smartwatches [61], Huawei watch [67] | https://fudickar-lab.github.io/WHIR/CodeSystem/attachment-method-cs | watch-strap | Watch strap | Device worn on the wrist via a watch-style strap, for example a smartwatch or wrist-mounted sensor, distinct from the general-purpose Strap concept reused from SNOMED CT. |
| Not observed in the reviewed corpus (Table S1); included for medical-grade adhesive sensor patches | https://fudickar-lab.github.io/WHIR/CodeSystem/attachment-method-cs | adhesive | Adhesive | Device fixed directly to the skin with a medical-grade adhesive film or patch, without any strap, clip, or garment involved. |
| Not observed in the reviewed corpus (Table S1); included for sensors sewn into or integrated with a garment | https://fudickar-lab.github.io/WHIR/CodeSystem/attachment-method-cs | garment-integrated | Garment-integrated | Sensor built into, sewn into, or otherwise integrated with a garment, so that the garment itself carries the sensor rather than holding a separate, removable device. |
| Free-form pocket or belt orientation [63] | https://fudickar-lab.github.io/WHIR/CodeSystem/attachment-method-cs | unconstrained-carry | Unconstrained carry | Device carried by the participant without a prescribed attachment or position, for example a smartphone kept wherever |
| | | | | the participant chose during the recording. Recorded explicitly so that a deliberately unconstrained protocol is distinguishable from an attachment method that is simply unknown or unreported. |
| Elastic straps [60] | https://fudickar-lab.github.io/WHIR/CodeSystem/attachment-method-cs | elastic-strap | Elastic strap | Device worn on the body via an elastic strap, distinct from the general-purpose Strap concept reused from SNOMED CT. |
| Upper-arm (mid-humerus) strap [5] | http://snomed.info/sct | 48990009 | Strap | General-purpose strap concept reused from SNOMED CT, for straps that are neither watch-style nor elastic. |

Table S4. Code systems and value sets defined by the proposed model, in addition to the profile- and extension-level artifacts of Table 3 and Table 4.

| Name | Kind | Concepts / source | Binding | Notes |
|---|---|---|---|---|
| PlacementQualifierCS | Code System | 2 locally defined concepts (preferred, affected) | — | Supplies the local codes used by the PlacementQualifier ValueSet (Table 5) |
| AttachmentMethodCS | Code System | 12 locally defined concepts | — | Supplies the local codes used by the AttachmentMethod ValueSet (Table S3) |
| HARActivityLabelsCS | Code System | 54 locally defined concepts | — | Supplies the codes used by the HARActivityLabels ValueSet |
| GeriatricAssessment LabelsCS | Code System | 34 locally defined concepts | — | Supplies the codes used by the GeriatricAssessmentLabels Value Set |
| ADLLabelsCS | Code System | 42 locally defined concepts | — | Supplies the codes used by the ADLLabels ValueSet |
| AnnotationProvenanceCS | Code System | 4 locally defined concepts (human-expert, human-reviewed, semi-automated, algorithm-generated) | — | Supplies the codes used by the AnnotationProvenance ValueSet |
| SensorModalityCS | CodeSystem | 6 locally defined concepts (accelerometer, linear acceleration, gravity, gyroscope, magnetometer, attitude) | — | Supplies the codes used by the SensorModality and IMUDeviceMetricType ValueSets |
| SensorAxisCS | CodeSystem | 4 locally defined concepts (x, y, z, magnitude) | — | Supplies the codes used by the SensorAxis ValueSet |
| DeviceIdentifierTypeCS | CodeSystem | 3 locally defined concepts (SYSID, BTMAC, ETHMAC) | — | Types of Device.identifier for sensor devices and gateways; same codes as ContinuaDeviceIdentifiers in the PHD IG, defined locally because that code system is only published for FHIR R4 |
| RawDataFileFormatCS | CodeSyste | 1 locally defined concept (whir-csv) | — | Format code of the raw data file, used in |
| | m | | | DocumentReference.content.profile |
| IMUSensorDevice Categories | ValueSet | 6 FHIR device-category codes | extensible | Device category for IMUSensorDevice |
| IMUSensorTypes | ValueSet | 2 SNOMED CT concepts and 1 ISO/IEEE 11073 concept | extensible | Sensor type for IMUSensorDevice |
| IMUDeviceMetricType | ValueSet | ISO/IEEE 11073 generic meter plus all SensorModalityCS concepts | extensible | Metric type for IMUDeviceMetric |
| HARBodyLocations | ValueSet | 39 SNOMED CT body-structure concepts (Section 2.3) | extensible | Table S2 (Supplementary File 2) lists the 33 codes used by the reviewed datasets |
| PlacementQualifier | ValueSet | 4 SNOMED CT concepts (dominant, nondominant, ipsilateral, contralateral) plus 2 locally defined (preferred, affected) | extensible | Full codes given in Table 5 |
| AttachmentMethod | ValueSet | 12 locally defined mechanisms plus 1 reused SNOMED CT concept (strap) | extensible | Full codes given in Table S3 |
| HARActivityLabels | ValueSet | 54 locally defined concepts | extensible | General-HAR use case |
| GeriatricAssessment Labels | ValueSet | 34 locally defined concepts and walking from HARActivityLabelsCS | extensible | Clinical-assessment/rehabilitation/fall/mobility-transition side of the home-geriatric-assessment use case |
| ADLLabels | ValueSet | 42 locally defined concepts | extensible | Passively observed activities-of-daily-living side of the same use case (Section 3.3) |
| ActivityLabels | ValueSet | Unions HARActivityLabels, GeriatricAssessment Labels, and ADLLabels | extensible | Umbrella binding for ActivityLabelObservation.code |
| FunctionalAssessmentTests | ValueSet | 2 LOINC clinical-test codes (TUG, 30-second chair stand test) | extensible | FunctionalAssessmentObservation.code; results of standardized functional tests in the home-geriatric-assessment use case |
| AnnotationProvenance | ValueSet | 4 locally defined concepts | required | Exhaustively enumerated concept, so bound with required rather than extensible strength (Section 3.3) |
| SensorModality | ValueSet | 6 locally defined concepts | extensible | Sensor modality in LongRawSensorModalityObservation.code |
| SensorAxis | ValueSet | 4 locally defined concepts | extensible | Axis-to-column mapping in LongRawSensorModalityObservation.component.code |
| IMUSensorUnits | ValueSet | 6 UCUM units | extensible | Unit of IMUDeviceMetric |